\documentclass[12pt]{article}

\usepackage{setspace}

\usepackage{amsmath}
\usepackage{amsfonts}
\usepackage{amssymb}
\usepackage[T1]{fontenc}
\usepackage{lmodern}
\usepackage{graphicx}
\usepackage{multirow}
\usepackage{multicol}
\usepackage{makecell}
\usepackage{booktabs}
\usepackage{tikz}
\usepackage{soul}
\usepackage{pgfplots}
\pgfplotsset{compat=1.17}
\usepackage{caption}
\usepackage{subcaption}
\usepackage{float}
\usepackage[section]{placeins}
\usepackage{comment}
\usepackage{url}
\usepackage[title]{appendix}

\newcommand{\Sigmav}{\mbox{\boldmath{$\Sigma$}}}

\newcommand{\piv}{\mbox{\boldmath{$\pi$}}}

\newcommand{\bI}{\mathbf{I}}

\newcommand{\by}{\mathbf{y}}

\newcommand{\bZ}{\mathbf{Z}}

\newcommand{\bmu}{\boldsymbol{\mu}}
\newcommand{\bSigma}{\boldsymbol{\Sigma}}
\newcommand{\real}{\text{\rm I\hspace{-0.6mm}R}} 
\newcommand{\tr}{\,\mbox{tr}}

\def\E{\mathbb E}

\def\Yv{\mathbf Y}
\def\Zv{\mathbf Z}

\def\Sv{\mathbf S}

\def\Wv{\mathbf W}

\def\xv{\mathbf x}
\def\yv{\mathbf y}

\def\1v{\mathbf 1}
\def\0v{\mathbf 0}

\def\mv{\mathbf m}

\newcommand{\vecd}{\mathbf{d}}
\newcommand{\diag}{\,\mbox{diag}}

\title{Model-based bi-clustering using multivariate Poisson-lognormal with general block-diagonal covariance matrix and its applications}

\author{Caitlin Kral\footnote{School of Mathematics and Statistics, Carleton University, Ottawa, ON, Canada}  \quad Evan Chance \footnote{School of Mathematics and Statistics, Carleton University, Ottawa, ON, Canada} \quad Ryan Browne\footnote{Department of Statistics and Actuarial Science, University of Waterloo, Waterloo, ON, Canada} \quad Sanjeena Subedi\footnote{School of Mathematics and Statistics, Carleton University, Ottawa, ON, Canada.}}

\begin{document}

\maketitle

\begin{abstract}
{While several Gaussian mixture models-based biclustering approaches currently exist in the literature for continuous data, approaches to handle discrete data have not been well researched. A multivariate Poisson-lognormal (MPLN) model-based bi-clustering approach that utilizes a block-diagonal covariance structure is introduced to allow for a more flexible structure of the covariance matrix. Two variations of the algorithm are developed where the number of column clusters: 1) are assumed equal across groups or 2) can vary across groups. Variational Gaussian approximation is utilized for parameter estimation, and information criteria are used for model selection. The proposed models are investigated in the context of clustering multivariate count data. Using simulated data the models display strong accuracy and computational efficiency and is applied to breast cancer RNA-sequence data from The Cancer Genome Atlas.}
\end{abstract}

\section{Introduction}

Bi-clustering is a technique that simultaneously clusters observations and features (i.e., variables) in a dataset \cite{Chakraborty2021}. This approach can be used in a bioinformatics setting to simultaneously identify clusters of diseased and non-diseased patients along with the groups of genes with distinct correlation patterns within each cluster. In addition, when applied to gene expression data, bi-clustering provides flexibility to identify co-expressed genes under some, but not all conditions, that the traditional clustering methods lack \cite{Liu}. Early gene expression studies relied on low-throughput methods such as northern blots and quantitative polymerase chain reaction (qPCR); however, these are limited to measuring single transcripts \cite{Kukurba}. More advanced methods were developed that enabled genome-wide quantification of gene expression which was performed by hybridization-based microarray technologies but many limitations existed \cite{Kukurba}. For example, for microarrays, prior knowledge of sequences is required; there can be problematic cross-hybridization artifacts in the analysis of highly similar sequences; there is a limited ability to accurately quantify lowly and very highly expressed genes \cite{Kukurba}. These restrictions motivated the development of high-throughput next-generation sequencing (NGS) by enabling the sequencing of complementary RNA (cRNA) \cite{Kukurba}. This method, known as RNA sequencing (RNA-seq), provides a more complete and quantitative view of gene expression, alternative splicing, and allele-specific expression \cite{Kukurba}. The increasing popularity of RNA-seq has prompted the fast-growing need for appropriate statistical models, bioinformatics expertise and computational resources to model these datasets efficiently \cite{Koch}. RNA-seq gene expression data is different from microarray data - they are discrete counts and they contain abundant zeros as not all genes are expressed under specific experimental conditions; therefore, algorithms that have been designed and evaluated with microarray data may not be suitable to apply to RNA-seq data \cite{Xie}. Only a few bi-clustering algorithms exist for RNA-seq data. As RNA-seq data becomes more popular, a knowledge gap for applying bi-clustering tools may soon become apparent \cite{Xie}.

Several bi-clustering algorithms have been developed in recent years. Padilha and Campello \cite{Padilha2017} examined 17 different biclustering algorithms selected from previous studies and references in the literature. They broadly classified these biclustering algorithms into four groups: Greedy algorithms, divide-and-conquer algorithms, exhaustive enumeration algorithms, and distribution parameter identification algorithms. Greedy algorithms are based on greedy approaches that hope to obtain a global optimal solution by performing the best local decision at each iteration. These include algorithms such as Cheng and Church's algorithm (CCA) \cite{Cheng2000}, QUalitative BIClustering (QUBIC) \cite{Li2009}, and Iterative Signature Algorithm (ISA) \cite{Kanehisa2000}. Divide-and-conquer approaches divide a larger problem into smaller components and then solve each component recursively. All the solutions are then combined into a single solution for the original problem with a primary example being the Binary Inclusion-Maximal Biclustering algorithm (Bimax)\cite{Prelic2006}. Exhaustive enumeration \cite{Madeira2004} assumes the most adequate submatrices can only be recognized by producing all potential row and column combinations of a dataset. By restricting the size of the searched biclusters these methods aim to avoid exponential running time and include algorithms such as the Statistical-Algorithmic Method for Bicluster Analysis (SAMBA) and Bit-Pattern Biclustering (BiBit) \cite{Rodriguez2011, Tanay2002}. Lastly, distribution parameter identification algorithms uses a statistical distribution with a parameter that is related to the structure of the biclusters and iteratively updates the parameter. Many algorithms fall within this sector including Plaid \cite{Turner2005}, Spectral \cite{Kluger2003}, Bayesian Biclustering (BBC) \cite{Gu2008}, and Factor Analysis for Bicluster Acquisition (FABIA) \cite{Hochreiter2010}. 

Challenging issues arise from both practical and theoretical perspectives such as dealing with potential heterogenous dimensions, missing data, outliers, as well as the intricacies of selecting correct representations \cite{Biernacki2023}. By employing a model-based clustering (MBC) framework, the assumption that a cluster is best represented by a specific probability distribution can be leveraged using both the power of mathematical statistics and mixture modeling flexibility \cite{Biernacki2023}. In the literature, bi-clustering in a finite mixture model framework has been proposed by imposing a block-diagonal covariance structure. The variables in the same blocks are assumed to belong to the same group and variables in different column clusters are believed to be uncorrelated. Other renditions of biclustering models also exist such as co-clustering. Co-clustering is a bi-clustering model which assumes all individuals belong to a singular row cluster and all variables belong to a singular column cluster \cite{Biernacki2023}. Unlike co-clustering, biclustering does not make this assumption as it aims to detect homogenous blocks within the data matrix which do not cover the entire matrix and which may overlap \cite{Biernacki2023}. While biclustering offers greater freedom in this respect, there tends to be more algorithmic complexity. Recently, Livochka et al. (2023) proposed a model-based biclustering framework that utilizes a mixture of multivariate Gaussian models with a block-diagonal covariance structure (comprising positive and negative correlations within the blocks) \cite{livochka2023}. Livochka et al. (2023) show that it produces comparable clustering performance to state-of-the-art algorithms while providing substantial improvement in the computational time \cite{livochka2023}.  

Multivariate count data is commonly encountered through high-throughput sequencing technologies in bioinformatics. Currently, several Gaussian mixtures model-based bi-clustering approaches exist for continuous data; however, approaches to handle multivariate count data have not been well researched. Previous work conducted by Martella et al. (2008) extended the Mixture Factor Analysis (MFA) model for a biclustering framework by replacing the component-specific factor loadings matrix with a binary row stochastic matrix, $\boldsymbol{B}$, which represented the column cluster membership. However, this model could only recover a restrictive covariance structure such that the off-diagonal elements in the block structure of the covariance matrices are restricted to be 1. Tu and Subedi \cite{Tu2022} expanded the work by Martella et al. (2008) by modifying the assumptions for the latent factors in the factor analyzer structure in order to capture a wider range of covariance structures. The loading matrix used in \cite{Martella2008} was replaced by a sparsity matrix. Under this assumption, the variables are clustered according to the underlying latent factors as each variable can only be represented by one factor and the variables represented by the same factors are clustered together \cite{Tu2022}. By assuming the diagonal covariance matrix to be a diagonal matrix, the component-specific covariance matrix becomes a block-diagonal matrix and within the block matrix, the off-diagonal elements are not restricted to 1; therefore, allowing for more flexible covariance structure recovery \cite{Tu2022}. Latent block models (LBM) have also been applied in within a model-based bi-clustering framework with adjustements to handle over-dispersed multivariate count data. For instance, Aubert and collegues \cite{Aubert2021} proposed a LBM with 
a negative binomial emission distribution. They adopted a parameterization of the Poisson– Gamma distribution which leads to a third hidden layer of the model that corresponds to a random effect that accounts for over-dispersion \cite{Aubert2021}. Additional advancements within bi-clustering algorithms has included the use of a hierarchical clustering being adopted within the clustering framework. Hierarchical clustering aims to generate a nested sequence of partitions incorporating them into a tree-like structure \cite{Grun2019}. Martella and collegues \cite{Martella2011} proposed a hierarchical extension to their model proposed in \cite{Martella2008} in order to allow for gene clusters to differ from mixture components and to identify blocks of genes and experimental conditions. Furthermore, by implementing a hierarchical structure, the parametric assumptions imposed upon the cluster-specific gene expressison distributions is avoided \cite{Martella2011}. The adopted hierarchical structure distinguished components from clusters and was able to more effecitvely cluster observations in comparison to the standard Gaussian mixture models \cite{Martella2011}. 

Recent literature has highlighted various different types of clustering validation criteria for hierarchical clustering frameworks. Cluster validation measures can be categorized into external, internal, and relative methods where the applicability of the methods is dependent on the problem at hand. Pagnuco et al. \cite{Pagnuco2017} presented a new algorithm that combined particularities of a hierarchical clustering algorithm for analyzing genetic association. Furthermore, more recent work by Gere (2023) also investigated hierarchical clustering validation in a consumer sensory project setting. When it comes to cluster validation, there is no golden standard as the results depend on the dataset \cite{Gere2023}. In addition, there are multiple validation methods available which may provide differing results \cite{Gere2023}. However, both works utilize one of the most popular validation criteria, the Silhouette index. The Silhouette index can explicitly ponderate intracluster's compactness versus intercluster distances and it can be adapted to measure the quality of the unique clusters instead of the overall clustering \cite{Pagnuco2017}. In addition, this index has also proven to be reliable when the cluster shape is not expected to be too extreme \cite{Pagnuco2017}. 

To model the multivariate count data arising from RNA-seq data, as well as bioinformatics in general, multivariate Poisson-lognormal (MLPN) models have been proposed in recent literature \cite{subedi2020}. Here, a model-based bi-clustering approach for RNA-seq data using the mixture of MPLN distribution that utilizes a block-diagonal covariance structure is proposed. 
 


\section{Methodology}

\subsection{The MPLN Model}

The MPLN distribution can be obtained using the hierarchical Poisson structure:

\begin{align*}
    Y_{ij}|X_{ij}&\sim \text{Poisson}\left(\exp\{X_{ij}\} \right)\quad \text{and}\quad  X_i\sim \mathcal{N}_d(\bmu, \bSigma),
\end{align*}

\noindent where ($X_{i1}, ..., X_{id}$) and $\mathcal{N}_d(\bmu, \bSigma)$ denotes a $d$-dimensional Gaussian distribution with mean $\boldsymbol{\mu}$ and covariance $\bSigma$. The resulting marginal distribution of $\mathbf{Y}$ is an MPLN distribution. Furthermore, working with RNA-seq data requires the incorporation of a normalization factor into the MPLN distribution \cite{Silva2019}. A normalization factor removes systematic technical effects that occur in data to ensure technical bias minimally impacts results \cite{Robinson2010}. Similar to \cite{Silva2019,Payne2023}, to account for the differences in library sizes across the samples, we incorporate a fixed known constant $C_i$ representing the normalized library sizes in the mean of the Poisson distribution such that the expected value of the abundance of the $j^{th}$ transcript from $i^{th}$ sample is $\E(Y_{ij}|X_{ij})=\exp\{X_{ij}+\log C_i\} $ .

A $G$-component mixture of MPLN~distributions can be written as: 
\begin{align*}
    f(\by\mid\boldsymbol{\vartheta}) &=\sum_{g=1}^G\pi_g~ f(\by\mid\bmu_g,\bSigma_g),
\end{align*}
where $\pi_{g}$ represents the mixing proportions, $\boldsymbol{\vartheta}$ denotes the model parameters, and $f_Y(\boldsymbol{y|\mu_g,\boldsymbol{\Sigma_g}})$ denotes the $g^{th}$ component MPLN distribution with parameters $\boldsymbol{\mu_g}$ and $\boldsymbol{\Sigma_g}$. 

For the clustering of observations (from here on referred to as row-clusters), we introduce a latent variable $\bZ$ such that $Z_{ig}=1$ if an observation belongs to group $g$ and 0 otherwise. In the clustering context, the $\bZ$ is treated as missing data and is estimated conditional on the observed data. Since bi-clustering aims to group both the columns and observations, grouping of the variables (from here on referred to as column-groups) is introduced by imposing a block structure on the covariance matrix $\bSigma_g$ with $K$ blocks where $K$ is the number of column-clusters. This structure implies that the variables in the same column-group have non-zero covariance and variables in different column-groups are assumed to be independent. 

\begin{figure}[!htb] 
\centering
    \includegraphics[width=0.6\textwidth]{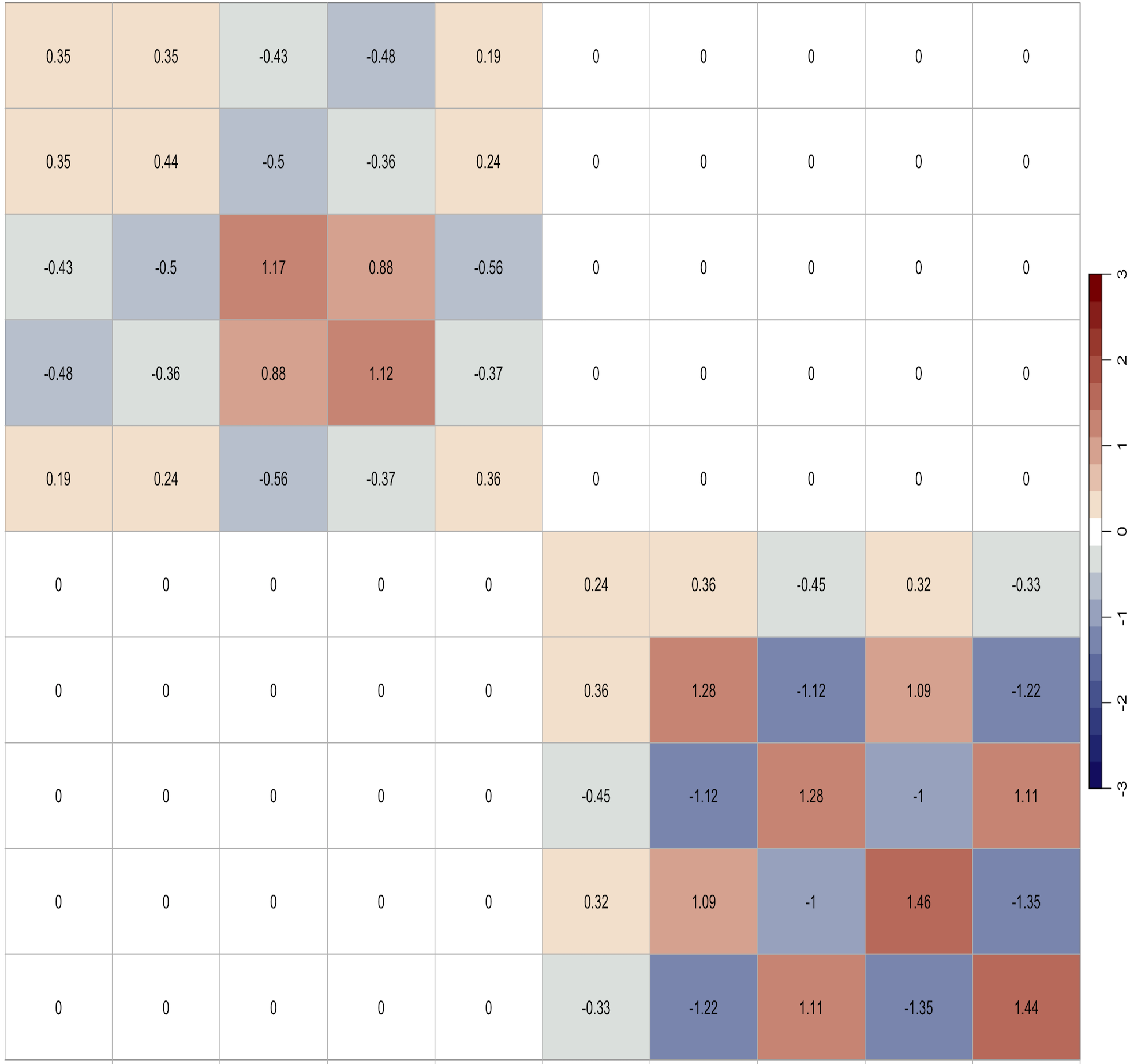}
    \caption{Illustration of block diagonal matrix with two column-groups.}
    \label{fig: block diag}
\end{figure}

Similar to \cite{livochka2023}, we propose to parameterize the block-diagonal covariance matrix as:

\begin{equation*}     
\label{eq:param}
    \bSigma_{g}= 
    \diag(\vecd_{g1})^{\intercal} \bSigma \diag(\vecd_{g1}) + \dots + \diag(\vecd_{gK})^{\intercal} \bSigma \diag(\vecd_{gK}),
\end{equation*}

where $\diag(\vecd_{g1}) +  \dots  + \diag(\vecd_{gK}) = \bI_p$,  and each element of $\vecd_{gk}$ is zero or one. The $\vecd_{gk}$ are binary vectors which allocate the variables to each block in the $g^{th}$ component. An example of the block-diagonal structure imposed on $\bSigma_g$ for biclustering is shown in Figure~\ref{fig: block diag}. 
  
Figure \ref{fig: block diag} represents a general block-diagonal structure that the covariance matrices for the underlying mixture components take on. Both positive and negative covariance values can be allowed leading to an unrestricted covariance structure. This is critical as multivariate count data, such as RNAseq data, are high-dimensional and genes involved in related biological pathways can exhibit positive or negative correlation among each other while genes involved in unrelated biological pathways may exhibit no correlation. Thus,  the addition of a block-diagonal covariance structure will also increase flexibility for modelling count data\cite{livochka2023}. As we can see for a set of the simulated data with a dimension of 10 and MLPN mixture of $G=2$, two blocks are formed. The blocks fall on the diagonal with non-zero covariance values and all remaining zero entries appear on the off-diagonal. 

\subsection{Parameter Estimation}

 Using the marginal probability mass function of observed data $\Yv$ and unobserved component indicator variable $\Zv$, the complete-data likelihood  can be defined as:
\begin{align*}
L(\boldsymbol{\vartheta}) &= \prod_{g=1}^G\prod_{i=1}^n\left[\pi_g f_Y(\yv|\bmu_g,\boldsymbol{\bSigma_g},C_i)~\right]^{z_{ig}},
\end{align*}
and the complete-data log-likelihood becomes:
\begin{align*}
l(\boldsymbol{\vartheta}) &= \sum_{g=1}^G\sum_{i=1}^n\left[z_{ig}\log \pi_g +\log f_Y(\mathbf{y}| \boldsymbol{\mu}_g, \bSigma_g,C_i ) \right]^{z_{ig}}.
\end{align*}

For the MPLN distribution, the marginal probability mass function of $\Yv$ is
\begin{align*}
 f(\by_i\mid \bmu_g,\bSigma_g)=\int_{\real^d} \left[ \prod _{j=1}^d f_p(y_{ij}\mid x_{ij},C_i)\right] ~f_N(\xv_i\mid\bmu_g,\bSigma_g)~d\xv_i,
\end{align*}
which involves multiple integrals and cannot be further simplified. Hence, similar to \cite{subedi2020}, we utilize variational Gaussian approximation (VGA) to approximate the $\log f_Y(\yv|\bmu_g,\boldsymbol{\Sigma_g})$. In VGA, a computationally convenient Gaussian distribution $q(\xv_{ig})\sim N(\mv_{ig},\Sv_{ig})$ is used to approximate the posterior density of $f(\xv_{i}\mid \yv_i,Z_{ig}=1)$ such that  the $q(\xv_{ig})\sim N(\mv_{ig},\Sv_{ig})$ minimizes the Kullback-Leibler (KL) divergence between $f(\xv_{i}\mid \yv_i,Z_{ig}=1)$ and approximating distribution $q(\xv_{ig})$ . Suppose, we have an approximating density $q(\xv_{ig})$, we can write 

\begin{align*}
\log f(\by_i) &= D_{KL}(q_{ig}\|f_{ig})+F(q_{ig},\by_i),
\end{align*}

where $D_{KL}(q_{ig}\|f_{ig})= \int_{\real^d} q(\xv_{ig}) \log \frac{q(\xv_{ig})}{f(\xv_{i}\mid\by_i,Z_{ig}=1)} d\xv_{ig}$ is the Kullback-Leibler (KL) divergence between $f(\xv_{i}\mid \yv_i,Z_{ig}=1)$ and approximating distribution $q(\xv_{ig})$, and 
$$F(q_{ig},\by_i)=\int_{\real^d} \left[\log f(\by_i,\xv_i\mid Z_{ig}=1) - \log q(\xv_{ig})\right] q(\xv_{ig})  d\xv_{ig},$$ is evidence lower bound (ELBO) for each observation $\by_i$. When $q(\xv_{ig})\sim N(\mv_{ig},\Sv_{ig})$, the ELBO for each observation $\by_i$ can be written as:

\begin{align*}
F(q_{ig},\by_i) &= -\frac{1}{2}(\mv_{ig}-\bmu_g)'\bSigma_g^{-1}(\mv_{ig}-\bmu_g)-\frac{1}{2} \tr(\bSigma_g^{-1}\Sv_{ig})+\frac{1}{2} \log |\Sv_{ig}| -\frac{1}{2} \log |\bSigma_g| + \frac{d}{2} \\
&+ \mv_{ig}'\by_i +\sum_{j=1}^d\left(\log C_i\right)y_{ij}-   \sum_{j=1}^d \left\{e^{\log C_i+m_{igj}+\frac{1}{2}S_{ig,jj}}+\log(y_{ij}!)\right\}.
\end{align*}

The variational parameters $\mv_{ig}$ and $\Sv_{ig}$ that maximize the ELBO will minimize the KL-diverence between $f(\xv_{i}\mid \yv_i,Z_{ig}=1)$ and $q(\xv_{ig})$. For details, see Subedi and Browne \cite{subedi2020}.

Thus, using the ELBO, the lower bound of the complete-data log-likelihood can be written as:

\begin{align*}
\tilde l(\boldsymbol{\vartheta}) &= \sum_{g=1}^G\sum_{i=1}^n\left[z_{ig}\log \pi_g +F(q_{ig},\by_i)~\right]^{z_{ig}}.
\end{align*}

Thus, parameter estimation is performed using a variational variant of the EM algorithm similar to the one proposed by Subedi and Browne \cite{subedi2020}. At the $t+1$ iteration,

\begin{enumerate}  
    \item[Step 1:] Conditional on variational parameters $\boldsymbol{m}_{ig}^{(t)}$, $\boldsymbol{S}_{ig}^{(t)}$ and on $\piv_g$, $\boldsymbol{\mu}_g^{(t)}$ and $\boldsymbol{\Sigma}_g^{(t)}$, similar to \cite{subedi2020}, the $E(Z_{ig})$ is approximated using $\hat{Z}_{ig}$:
    \begin{align*}
    \hat{Z}_{ig}^{(t+1)} = \frac{\piv_g^{(t)}\exp[F(q_{ig}, \by_i)]}{\sum\limits_{h=1}^G\piv_h^{(t)}\exp[F(q_{ih}, \by_i)])}.
    \end{align*} 
    
    \noindent 

    \item[Step 2:] Estimates of the variational parameters $\Sv_{ig}$ and $\mv_{ig}$ are obtained by maximizing the lower bound of the complete-data log-likelihood. For updating the variational parameters $\Sv_{ig}$ and $\mv_{ig}$, a ``fixed-point''\cite{subedi2020} method and Newton's step is used respectively as the maximum likelihood approach does not yield closed-form expressions. Conditional on $\hat{Z}_{ig}^{(t+1)}$, $\boldsymbol{\mu}_g^{(t)}$ and $\boldsymbol{\Sigma}_g^{(t)}$, the updates of $\Sv_{ig}$ and $\mv_{ig}$ as shown below:
    
    \begin{align*}
    \Sv_{ig}^{(t+1)} =& \left[\Sigmav_g^{-1(t)} + \text{exp}\left[\mv_{ig}^{(t)} + \frac{1}{2}\text{diag}(\Sv_{ig}^{(t)})\1v_d^{T} \right]\right]^{-1},\\
    \mv_{ig}^{(t+1)} =& \mv_{ig}^{(t)} - \Sv_{ig}^{(t+1)}\left(\text{exp}\left[\mv_{ig}^{(t)} + \frac{1}{2}\text{diag}(\Sv_{ig}^{(t+1)})\right] + \Sigmav_g^{-1(t)}(\mv_{ig}^{(t)} - \bmu_g^{(t)}) - \by_i\right).
    \end{align*}

    \item[Step 3:] The updated estimates for $\piv_g$, $\bmu_g$, and $\bSigma_g$ are obtained using a maximum likelihood approach. The updates for $\piv_g$ and $\bmu_g$ are given as 
    
    \begin{align*}
    \hat{\piv}_g^{(t+1)} = \frac{\sum_{i=1}^n\hat{\bZ}_{ig}^{(t+1)}}{n},\quad \text{and}\quad \hat{\bmu}_g^{(t+1)} = \frac{\sum_{i=1}^n \hat{\bZ}_{ig}^{(t+1)} \mv_{ig}^{(t+1)}}{\sum_{i=1}^n \hat{\bZ}_{ig}^{(t+1)}}.\\
    \end{align*}
\end{enumerate}
   
To obtain the estimate for the block diagonal $\bSigma_g^{(t+1)}$, we implement a hierarchical clustering framework. Hierarchical clustering refers to a popular family of clustering algorithms \cite{Murtagh2014, Beeferman2000, Erica2018} that follow an iterative procedure for merging or splitting nested clusters. Merging or splitting is also known as bottom-up and top-down approaches respectively. Similar to Livochka et al.\cite{livochka2023}, the bottom-up hierarchical approach known as agglomerative clustering was utilized to find column-groups. This procedure starts by assigning every variable to its own group. During every iteration, it merges two of the most similar groups, as measured by a specific similarity/dissimilarity metric known as the linkage criteria. The linkage criteria is the metric that determines the distance between two clusters. Once the $K$ variable groups are determined, the corresponding block-diagonal covariance matrix can be constructed. To begin, we first need the estimates of $\vecd_1$,\ldots,$\vecd_K$. We first compute $\widehat{\Wv}_{g}^{(t+1)}$ which can be regarded as the unrestricted sample covariance matrix: 

\begin{align*}
    \widehat{\Wv}_{g}^{(t+1)} =& \frac{\sum_{i=1}^n\hat{\bZ}_{ig}^{(t+1)}(\mv_{ig}^{(t+1)}-\hat{\bmu}_g^{(t+1)})
    (\mv_{ig}-\hat{\bmu}_g^{(t+1)})}{\sum_{i=1}^n\hat{\bZ}_{ig}^{(t+1)}} + \frac{\sum_{i=1}^n \hat{\bZ}_{ig}^{(t+1)}\mathbf{S}_{ig}^{(t+1)}}{\sum_{i=1}^n\hat{\bZ}_{ig}^{(t+1)}}.
\end{align*}

 
The $\widehat{\Wv}_{g}^{(t+1)}$ is then converted into a sample correlation matrix and using these correlations, for each component $g$, we define the distance between the two variables $x_i$ and $x_j$ as $$\text{dist}_g(x_i,x_j) = 1 - \text{corr}(x_i,x_j)^2.$$ 

The distance measures represent dissimilarities between entries which can then be passed to the hierarchical clustering to determine column groupings. For each component $g$, the column groupings from the hierarchical clustering are then used to define a $d\times K$ column cluster membership indicator variable $\mathbf{D}_g$ such that 
 
	\[D_{g,jk} = \left\{\begin{array}{ll}
	1 & \text{if  variable}~x_j \text{ belongs to group } k,\\
	0 & \text{otherwise}.\\
	\end{array}\right.
	\]
where $k=1,\dots,K$ and the $k^{th}$ column of the $\mathbf{D}_g$ is such that $\mathbf{D}_g=[\vecd_{g1}$,\ldots,$\vecd_{gK}].$

Given $\boldsymbol{D}_g$, the estimate of the block-diagonal covariance matrix for the $g^{th}$ component with $K$ blocks is obtained as

\begin{align*}
    \hat{\boldsymbol{\Sigma}}_g^{(t+1)} =& \sum_{k = 1}^K \diag(\vecd_k)\widehat{\Wv}_{g}\diag(\vecd_k).
\end{align*}

We implement two different approaches for estimating the column clusters $K$. In the first approach, we assume the same number of $K$ blocks for each group. Then, for a range of $G$ and $K$ values, the model is fitted for all possible combinations of the different values of $G$ and $K$. The best-fitting model is identified $a$-posteriori using model selection criteria. The Bayesian information criterion (BIC)\cite{schwarz1978}, defined as:
$$ \text{BIC}=2l(\boldsymbol{\vartheta})-p \log n,$$
remains the most widely used model selection criteria for mixture models, where $l(\boldsymbol{\vartheta})$ is the complete-data log-likelihood, $p$ is the number of free parameters in the model, and $n$ is the sample size. In our case, we computed an approximation of BIC using the lower bound of the complete-data log-likelihood.

However, our assumption that $K$ is equal across groups is quite restrictive, especially within real data, as it presumes every group will have the same number of column-groups. While the algorithm does allow the flexibility of having different combinations of variables between different groups for the block structures in the covariance matrices, the number of blocks is restricted to be the same. Therefore, the data may be clustered in a manner that does not identify the best correlation patterns of each group. Thus, in the second approach, we set the maximum number of column-groups $K_{max}$ to be the same for all groups. Then, for each group, $g$, the algorithm is fitted for all possible values of $K$ ranging from 1 to $K_{max}$ and a cluster validation criterion is used for selecting the optimal $K_g$ for each group $g$.

We utilized the average Silhouette to determine the number of $K$ blocks for each group. The Silhouette index measures the quality of the clusterings as the average quality of its elements \cite{Rousseeuw1987}. The Silhouette index is defined by:



\begin{align*}
    S(X) =  \frac{b(X) - a(X)}{max(a(X), b(X))},
\end{align*}
where $a(X)$ is the average distance of point $X$ to the points of cluster $C_{k}$ and $b(X)$ is the average distance to the nearest cluster for a point $X \in C_{k}$. The Silhouette index for a cluster is then the average Silhouette of its points: 

\begin{align*}
    S(C_{k}) =  \frac{1}{n_{k}}\sum_{X \in C_{k}}S(X)
\end{align*}
where the value ranges from -1 to 1 such that a high Silhouette indicates the element is closer to its own cluster elements than the ones that do not belong to its own cluster. In order to select K, we first compute $S(C_{k})$ for all different $K$ and then select the K value associated with the largest average silhouette. 
To assess clustering performance, adjusted Rand index (ARI)\cite{hubert1985} is used. ARI measures the agreement between two partitions of data - ARI of 1 indicates perfect agreement and the expected value of ARI under random classification is 0. 

\subsection{Comparative Analysis}
To demonstrate the efficiency of the proposed biclustering framework, two other model-based biclustering algorithms were applied to the simulated data where the true underlying cluster structures are known for comparison. The first selected algorithm was the Gamma-Poisson Latent Block Model (PG-LBM) developed by Aubert and colleagues \cite{Aubert2021} and was utilized using the \texttt{cobiclust} package in \texttt{R}. This algorithm was selected for comparison as it is a model-based bi-clustering approach designed for over-dispersed count data and is based on partitioning the data matrix can be represented as $G\times K$ blocks where the data within the same blocks are modelled by the same parametric density. The second selected algorithm was developed by Tu and Subedi \cite{Tu2022} that employed an unrestricted Gaussian mixture model-based bi-clustering model (BMM), which was applied using the \texttt{bmm} package in \texttt{R}. Gaussian model-based approaches are a popular approach often cited in literature but this method is usually not ideal when handling over-dispersed count data. The BMM, however, is similar to the proposed method, such that it captures a wider range of covariance structures done through modification of the latent factor assumptions within the factor analyzer structure \cite{Tu2022}. Thus, the BMM is applied to the log-transformed counts. Comparing a Gaussian approach to the MPLN will highlight the robustness of both distributions when it comes to biclustering RNA-seq data. 

\section{Simulation Studies}

Simulation studies were completed in \texttt{R} statistical software \cite{R} with data simulated using the \texttt{mvtnorm} package \cite{mvtnorm}. In our simulations, the normalization constant $C_i$ is set to 1 as this data is not exposed to potential RNA-seq technical bias.

\subsection{Simulation Setting 1: $K$ is assumed to be equal across groups}
 In this simulation setting, we conducted ten sets of simulation studies. For each of the ten simulation studies, we generated 100 datasets of each of size $N$ from a $G$-component MPLN distribution with mixing proportions with $K$ different column-groups. For two-component models, the mixing proportions were set as $\piv = (0.25, 0.75)$ and for three-component models, the mixing proportions were set as $\piv = (0.35, 0.10, 0.55)$.  For all the simulation studies in the Simulation Setting 1, we fitted the models with $G=1,\ldots,G_{max}$ and $K=1,\ldots,K_{max}$ where $G_{max}$ was set to $\text{true value of G}+1$ and $K_{max}$ was set to $\text{true value of K}+1$.

In all ten simulation studies, BIC selected the correct model in all 100 out of 100 datasets. Furthermore, the selected value of $K$ matched the true value in 100 out of 100 datasets - indicating the correct model was selected 100\% of the time. The average ARI for row cluster recovery using the models selected by BIC was 1.00 with a standard deviation (sd) of 0.00. The average misclassification rates for the column-groups were 0\% in all simulation settings. Summary results of all ten simulation studies for Simulation Setting 1 are provided in Table~\ref{comp result}. 

\begin{figure}[!htb]
\centering
    \includegraphics[width=0.68\textwidth]{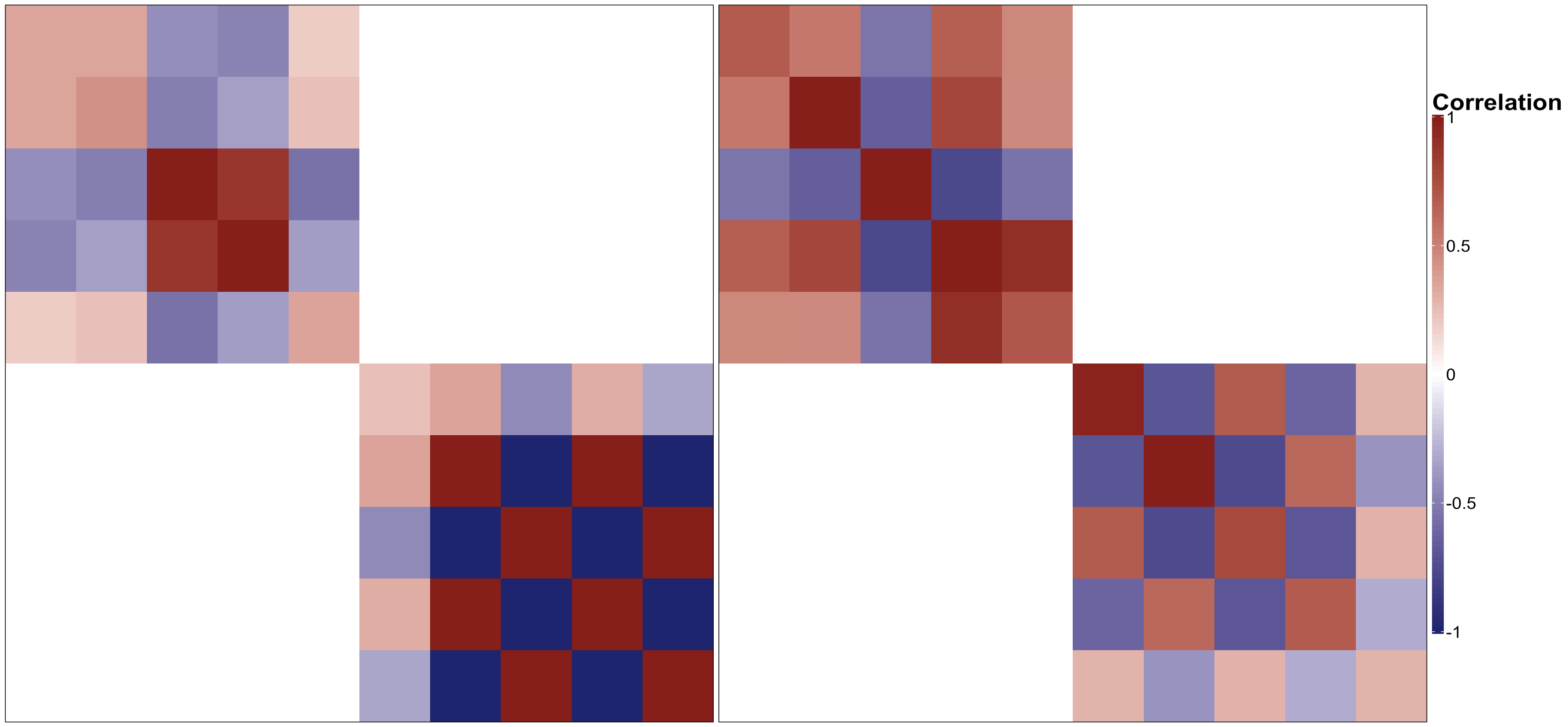}
    \includegraphics[width=0.68\textwidth]{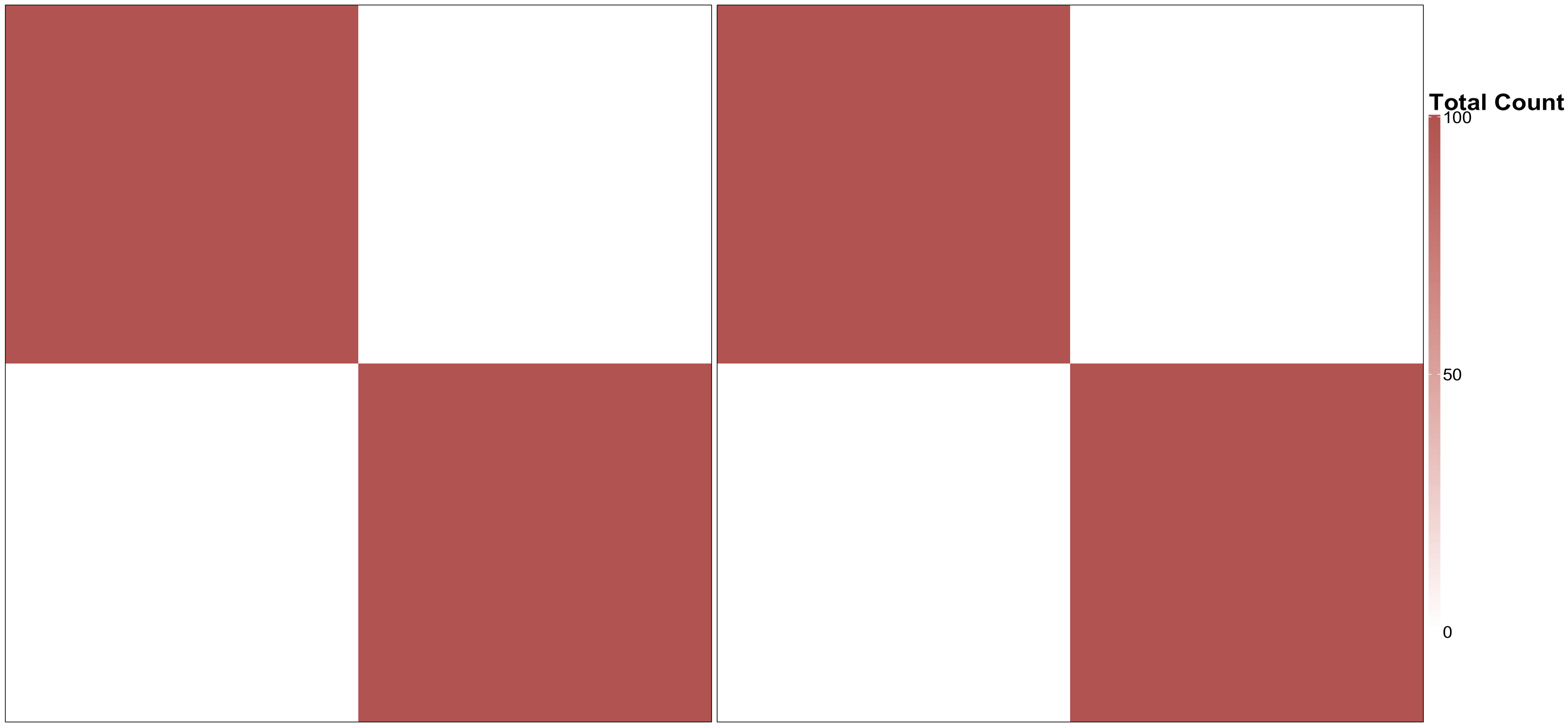}
    \caption{Covaraince matrices from generated data and the heatmap of the count for a non-zero entry within the covariance matrix for simlations of $N = 500$ given $d = 10$ with $G = 2$ and $K = 2$, respectively.}
    \label{fig:N500 cov matrices1}
\end{figure}

To assess the recovery of the column-groups and the corresponding covariance structure, we computed the number of times an entry within the covariance matrix was identified as non-zero for each entry in the covariance matrix. The heatmap of the number of times an entry in the two covariance matrices from simulation study 1 (with $N = 500$, $d=10$, $G=2$, and $K=2$) is provided in Figure \ref{fig:N500 cov matrices1}.

\begin{figure}[!htb]
\centering
    \includegraphics[width=0.68\textwidth]{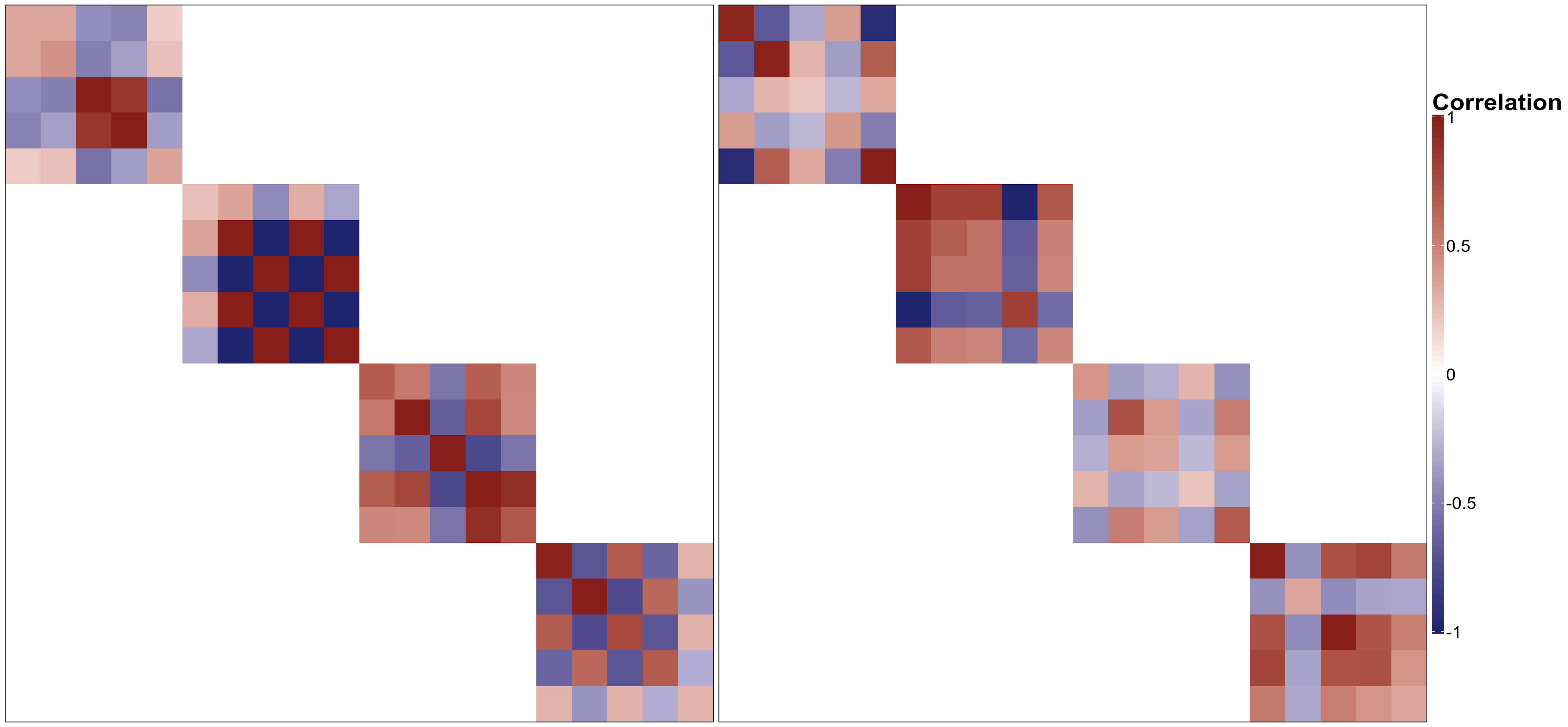}\\
    \includegraphics[width=0.68\textwidth]{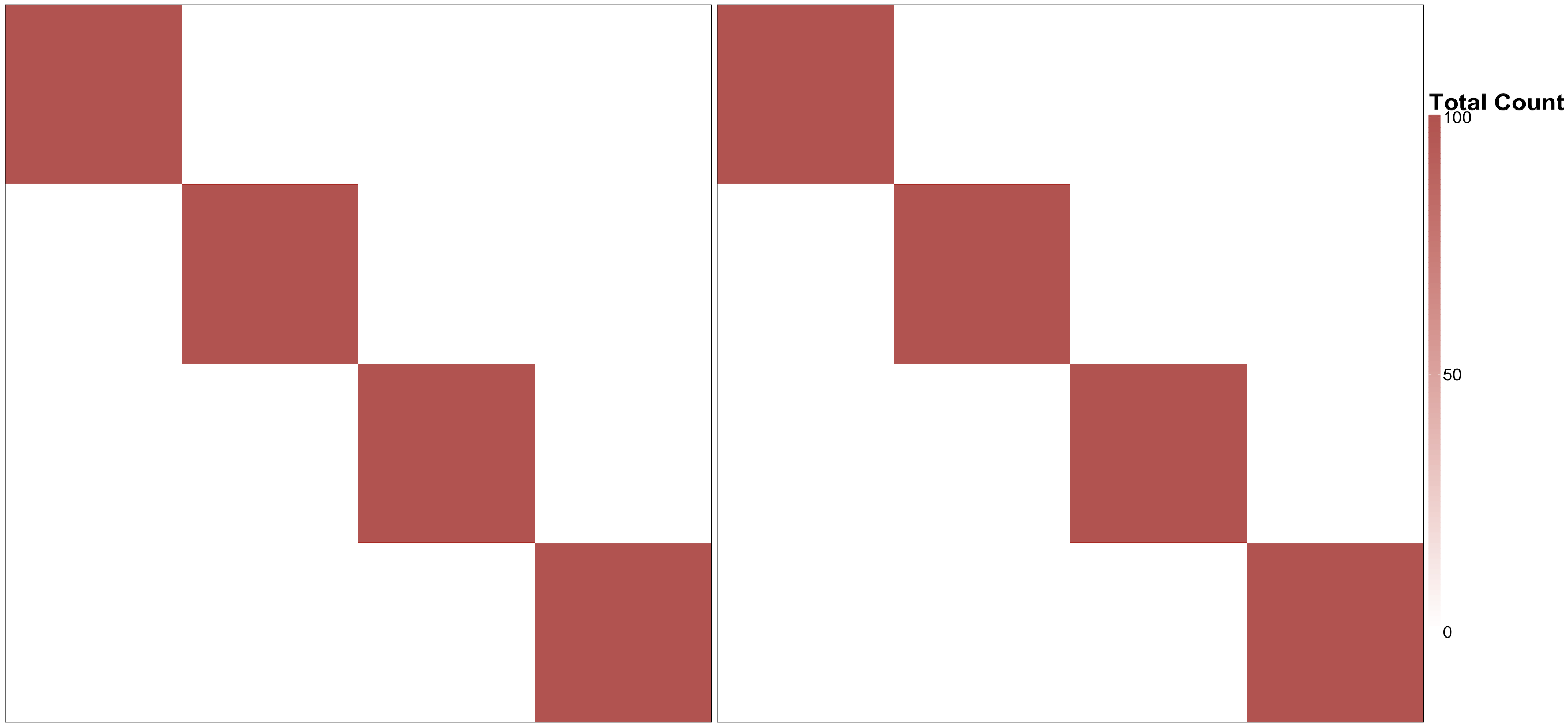}
    \caption{Covaraince matrices from generated data and the heatmap of the count for a non-zero entry within the covariance matrix for simlations of  $N = 500$ given $d = 20$ with $G = 2$ and $K = 4$, respectively.}\label{fig:N500d20}
    \label{fig:N500 cov matrices2}
\end{figure}

Perfect recovery of the covariance matrices is observed, with a perfect block formation occurring on the diagonal where the entries in the covariance matrices were non-zero. A similar trend was observed for the heatmap of the number of times an entry in the two covariance matrices from simulation study 2 (with $N = 500$, $d=20$, $G=2$, and $K=4$) as can be seen in Figure~\ref{fig:N500d20}.

\begin{center}
\begin{table}[!htb]
 \caption{Summary of the clustering performance in all 10 simulation studies from Simulation Setting 1 along with the comparison with PG-LBM and BMM.} \label{sim1} 
\scalebox{0.6}{
\begin{tabular}{@{\extracolsep{4pt}}cccccccc@{}}
\cline{1-8}
\\
N&\multicolumn{2}{c}{Model}&\multirow{2}{8em}{\centering True G \\ and True K}&\multirow{2}{8em}{\centering G Selected \\ (\# of times)}&\multirow{2}{10em}{\centering K Selected \\(\# of times)}&\multirow{2}{8em}{average ARI (sd)\\ for row clusters}&\multirow{3}{10em}{\centering mean \% misclassification (sd)\\ for column-groups}\\
\\
\\
\cline{1-8}\\

n = 500&\multicolumn{2}{c}{\textbf{Simulation study 1}}\\
&&Proposed&G = 2; K = 2&G = 2 (100)& K = 2 (100)&1.0 (0.00)&0.00\% (0.00\%)\\
&d = 10&PG-LBM&&G = 2 (100)& K = 2 (100)&0.97 (0.00)&50.00\% (0.00\%)\\
&&BMM&&G = 2 (100)&K = 3 (100)&1.0 (0.00)&15.00\% (5.00\%)\\[10pt]

&\multicolumn{2}{c}{\textbf{Simulation study 2}}\\
&&Proposed&G = 2; K = 4&G = 2 (100)&K = 4 (100)&1.0 (0.00)&0.00\% (0.00\%)\\
&d = 20&PG-LBM&&G = 2 (100)& K = 2 (100)&1.0 (0.00)&75.00\% (0.00\%)\\
&&BMM&&G = 3 (100)&K = 3 (100)&0.47 (0.00)&35.00\% (0.00\%)\\[10pt]

&\multicolumn{2}{c}{\textbf{Simulation study 3}}\\
&&Proposed&G = 2; K = 9&G = 2 (100)& K = 9 (100)&1.0 (0.00)&0.00\% (0.00\%)\\
&d = 50&PG-LBM&&G = 2 (100)&K = 10 (100)&1.0 (0.00)&62.00\% (0.00\%)\\
&&BMM&&G = 2 (100)&K = 7 (100)&1.0 (0.00)&49.00\% (3.00\%)\\[10pt]

&\multicolumn{2}{c}{\textbf{Simulation study 4}}\\ 
&&Proposed&G = 2; K = 10&G = 2 (100)& K = 10 (100)&1.0 (0.00)&0.00\% (0.00\%)\\
&d = 50&PG-LBM&&G = 2 (100)& K = 9 (100)&1.0 (0.00)&74.00\% (0.00\%)\\
&&BMM&&G = 2 (100)&K = 8 (100)&1.0 (0.00)&52.00\% (0.00\%)\\[10pt]

&\multicolumn{2}{c}{\textbf{Simulation study 5}}\\ 
&&Proposed&G = 2; K = 12&G = 2 (100)& K = 12 (100)&1.0 (0.00)&0.00\% (0.00\%)\\
&d = 50&PG-LBM&&G = 3 (100)& K = 12 (100)&0.50 (0.00)&64.00\% (0.00\%)\\
&&BMM&&G = 2 (100)&K = 6 (100)&1.0 (0.00)&67.00\% (3.00\%)\\[10pt]

n = 1000&\multicolumn{2}{c}{\textbf{Simulation study 6}}\\
&&Proposed&G = 3; K = 2&G = 3 (100)& K = 2 (100)&1.0 (0.00)&0.00\% (0.00\%)\\
&d = 10&PG-LBM&&G = 2 (100)&K = 3 (100)&0.83 (0.00)&30.00\% (0.00\%)\\
&&BMM&&G = 3 (100)&K = 3 (100)&1.0 (0.00)&13.00\% (5.00\%)\\[10pt]

&\multicolumn{2}{c}{\textbf{Simulation study 7}}\\
&&Proposed&G = 3; K = 4&G = 3 (100)& K = 4 (100)&1.0 (0.00)&0.00\% (0.00\%)\\
&d = 20&PG-LBM&&G = 4 (100)&K = 5 (100)&0.70 (0.00)&55.00\% (0.00\%)\\
&&BMM&&G = 4 (100)&K = 3 (100)&0.70 (0.00)&52.00\% (2.00\%)\\[10pt]

&\multicolumn{2}{c}{\textbf{Simulation study 8}}\\  
&&Proposed&G = 3; K = 9&G = 3 (100)& K = 9 (100)&1.0 (0.00)&0.00\% (0.00\%)\\
&d = 50&PG-LBM&&G = 4 (100)& K = 9 (100)&0.66 (0.00)&60.00\% (0.00\%)\\
&&BMM&&G = 1 (100)&K = 7 (100)&0.00 (0.00)&78.00\% (0.00\%)\\[10pt]

&\multicolumn{2}{c}{\textbf{Simulation study 9}}\\
&&Proposed&G = 3; K = 10&G = 3 (100)& K = 10 (100)&1.0 (0.00)&0.00\% (0.00\%)\\
&d = 50&PG-LBM&&GG = 4 (100)&K = 11 (100)&0.70 (0.00)&64.00\% (0.00\%)\\
&&BMM&&G = 4 (100)&K = 2 (100)&0.69 (0.00)&83.00\% (1.00\%)\\[10pt]

&\multicolumn{2}{c}{\textbf{Simulation study 10}}\\ 
&&Proposed&G = 3; K = 12&G = 3 (100)& K = 12 (100)&1.0 (0.00)&0.00\% (0.00\%)\\
&d = 50&PG-LBM&&G = 4 (100)& K = 11 (100)&0.67 (0.00)&66.00\% (0.00\%)\\
&&BMM&&G = 1 (100)&K = 10 (100)&0.00 (0.00)&70.00\% (0.00\%)\\[10pt]

\cline{1-8}
\end{tabular}
}
\end{table}
\label{comp result}
\end{center}

Visualization of the heatmap of the dataset with $d = 50$ for the larger sample size of $N = 1000$ are provided in Appendix~\ref{vis extra} in \ref{fig: N1000 d50 cov matrices1}. Even with larger sample sizes, the algorithm performs exceptionally with perfect recovery of all three covariance matrices - even when blocks are not symmetrical in size. Details of the parameter recovery for all ten simulation studies are provided in the Appendix~\ref{para}.

The proposed models were compared to the PG-LBM and BMM on all ten simulation studies. The BIC was used for model selection for all algorithms. Summary results of the number of times the correct row-cluster and column-group values were selected and the average ARI for row cluster and average misclassification along with their standard deviations for each method are shown in Table~\ref{sim1}. 
PG-LBM presents high ARI values for simulation studies 1-4 as it correctly selects the row-cluster value for all 100 datasets and fails to identify the column groups (with the average percentage misclassification $>$ 50\%). 
For simulation study 5, PG-LBM suffers in terms of both recovery of the row cluster and column groups.
Through simulation study 6-10, both PG-LBM and BMM see lower ARI values compared to the proposed method but PG-LBM consistently outperforms BMM. Across studies 6-10, PG-LBM has moderate to strong ARI values for row clusters with an average ARI of 0.71 while BMM suffers for simulation studies 8 and 10 highlighting the Gaussian mixture model's inability to handle over-dispered multivariate count data even after log-transformation. In simulations 3, 5, 8 and 10 block sizes are asymmetrical. PG-LBM still produces high ARI in simulation study 3 and moderate ARI in simulations 8 and 10. BMM produces strong ARI for simulations 3 and 5; however, simulations 8 and 10 both observe an ARI value of 0.00 for the row clusters. When examining the column group misclassification, both algorithms fail to outperform the proposed approach. 


\subsection{Simulation Setting 2: $K$ can vary across groups}
In this set of simulations, we generated 100 datasets of size $N = 500$ where the underlying latent variable \textbf{Y} came from two-component MPLN distribution with mixing proportions $\piv = (0.25, 0.75)$. In first simulation study in Simulation Setting 2 (i.e., Simulation Study 11), 100 ten-dimensional datasets were generated from a two-component MPLN and the number of column-groups $K_1$ and $K_2$ were set to 2 and 3 for $G=1$ and $G=2$ components, respectively. In the second simulation study in Simulation Setting 2 (i.e., Simulation Study 12), 100 twenty-dimensional datasets were generated from a two-component MPLN and the number of column-groups $K_1$ and $K_2$ were set to 4 and 5 for $G=1$ and $G=2$ components, respectively. Summary of the results for the two simulation studies from Simulation Setting 2 are provided in Table~\ref{tab:sim2} along with the sample size $N$, dimension of the dataset $d$, the true number of row-clusters $G$, and the true number of column-groups $K_1$ and $K_2$. 

\begin{center}
\begin{table}[!ht]
 \caption{Summary of the clustering performance in both simulation studies from Simulation Setting 2 in comparison with PG-LBM and BMM. Here, $N=500$ was used for both datasets.}\label{tab:sim2} 
\scalebox{0.58}{
\begin{tabular}{@{\extracolsep{4pt}}ccccccc@{}}
\cline{1-7}
\\
\multicolumn{2}{c}{Model}&\multirow{2}{8em}{\centering True G \\ and True K}&\multirow{2}{8em}{\centering G Selected \\ (\# of times)}&\multirow{2}{10em}{\centering K Selected \\(\# of times)}&\multirow{2}{8em}{average ARI (sd)\\ for row clusters}&\multirow{2}{10em}{\centering mean \% misclassification (sd)\\ for column-groups}\\
\\
\\
\cline{1-7}\\
\multicolumn{2}{c}{\textbf{Simulation study 11}}\\
&Proposed&\multirow{2}{8em}{\centering G = 2 \\ $K_1$ = 2;$ K_2$ = 3}&G = 2 (100)&$K_1$ = 2; $K_2$ = 3 (100)&1.0 (0.00)&0.00\% (0.00\%)\\
d = 10&PG-LBM&&G = 3 (100)&$K_1$ = 4; $K_2$ = 4 (100)&0.49 (0.00)&30.00\% (0.00\%)\\
&BMM&& G = 3 (100)&$K_1$ = 1; $K_2$ = 2; $K_3$ = 3 (100)&0.53 (0.00)&45.00\% (15.00\%)\\[5pt]
\multicolumn{2}{c}{\textbf{Simulation study 12}}\\
&Proposed&\multirow{2}{14em}{\centering G = 2 \\ $K_1$ = 4; $K_2$ = 5}&G = 2 (100)&$K_1$ = 4; $K_2$ = 5 (100)&1.0 (0.00)&0.00\% (0.00\%)\\
d = 20&PG-LBM&&G = 2 (100)&$K_1$ = 6; $K_2$ = 6 (100)&0.99 (0.00)&48.00\% (3.00\%)\\
&BMM&& G = 3 (100)&$K_1$ = 4; $K_2$ = 1; $K_3$ = 5 (100)&0.48 (0.00)&65.00\% (15.00\%)\\
\cline{1-7}
\end{tabular}
}
\end{table}
\label{comp vary result}
\end{center}

\begin{figure}[H] 
\centering
    \includegraphics[width=0.7\textwidth]{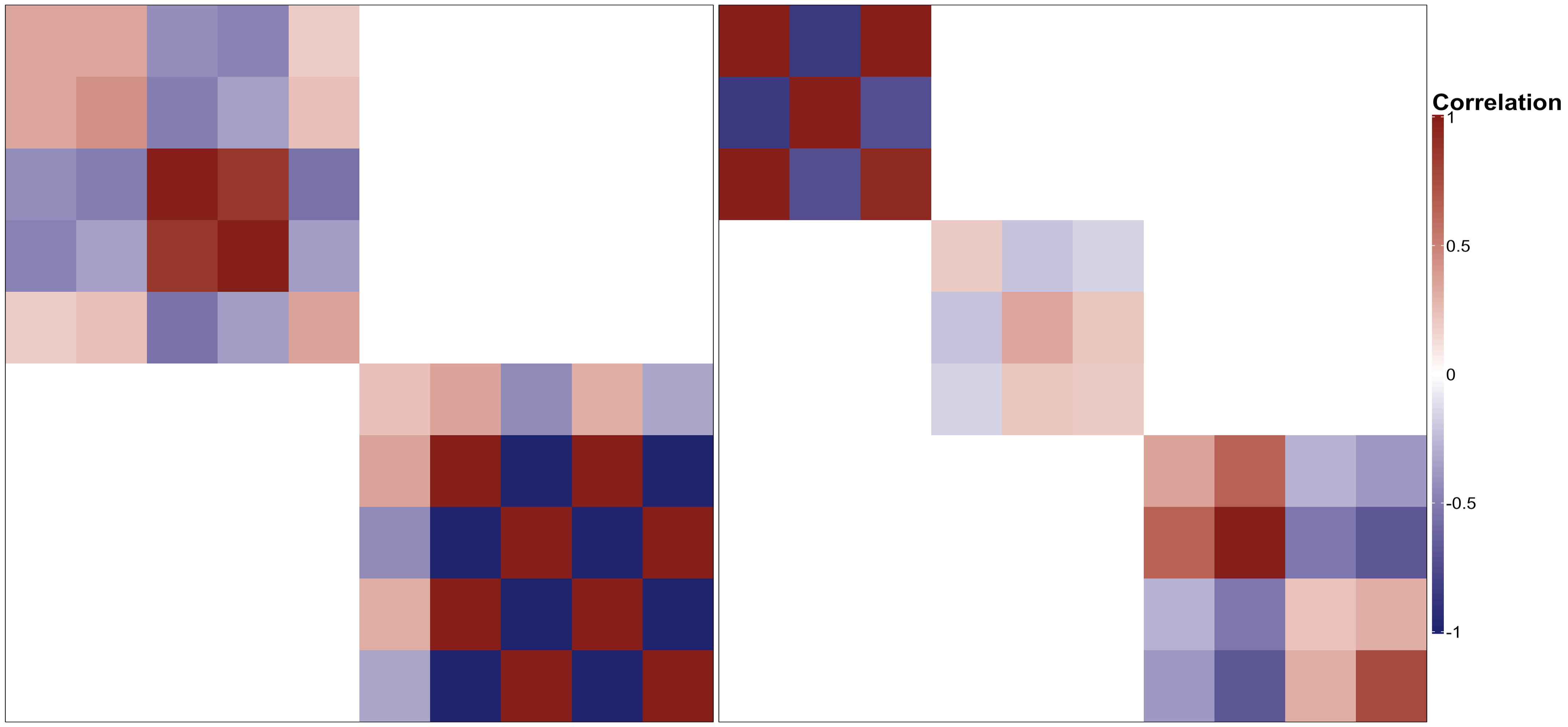}\\
    \includegraphics[width=0.68\textwidth]{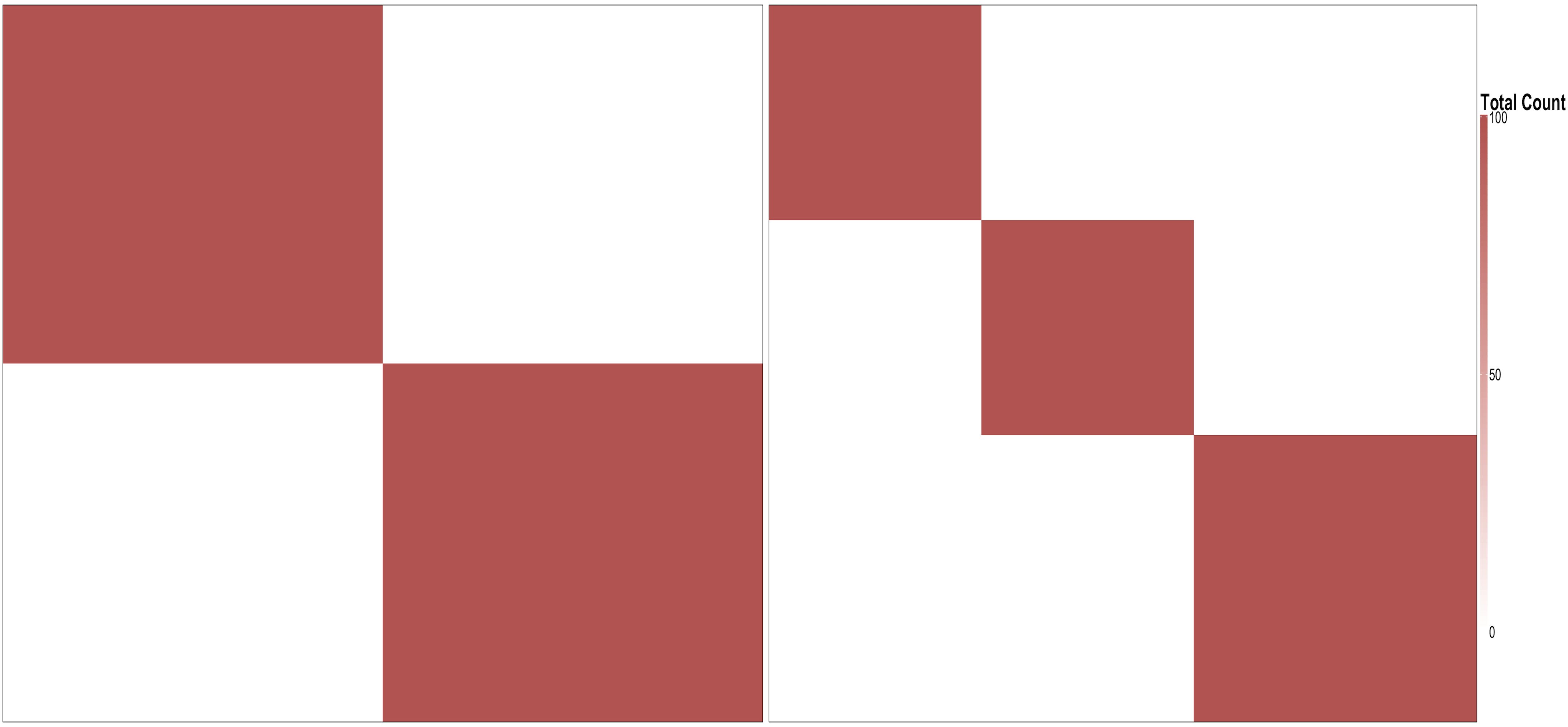}
    \caption{Covaraince matrices from generated data and the heatmap of the count for a non-zero entry within the covariance matrix for simulations of $N = 500$ given $d = 20$ with $G = 3$ and $K = 4$; $5$ respectively.}
    \label{fig:Vary d10 cov matrices}
\end{figure}

\begin{figure}[!htb] 
\centering
    \includegraphics[width=0.7\textwidth]{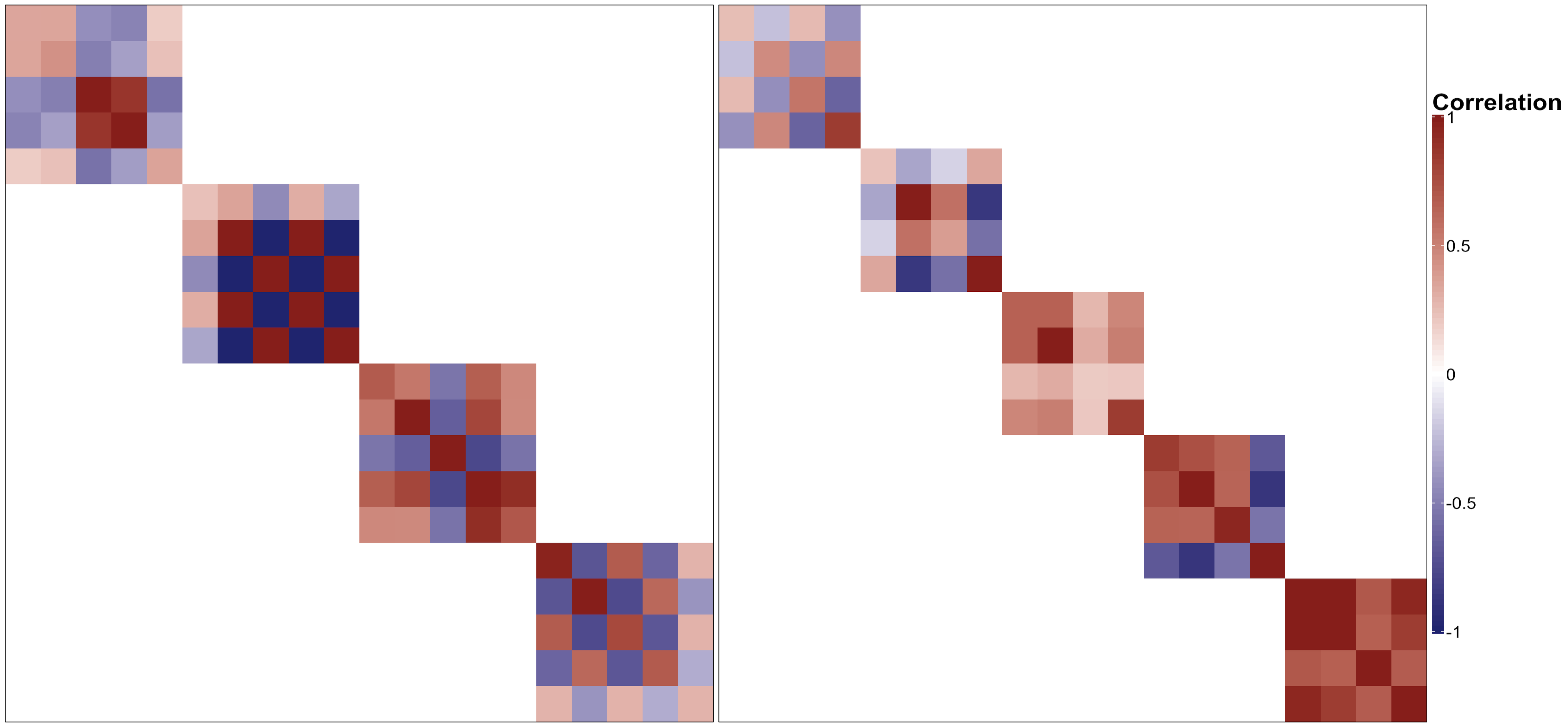}\\
    \includegraphics[width=0.68\textwidth]{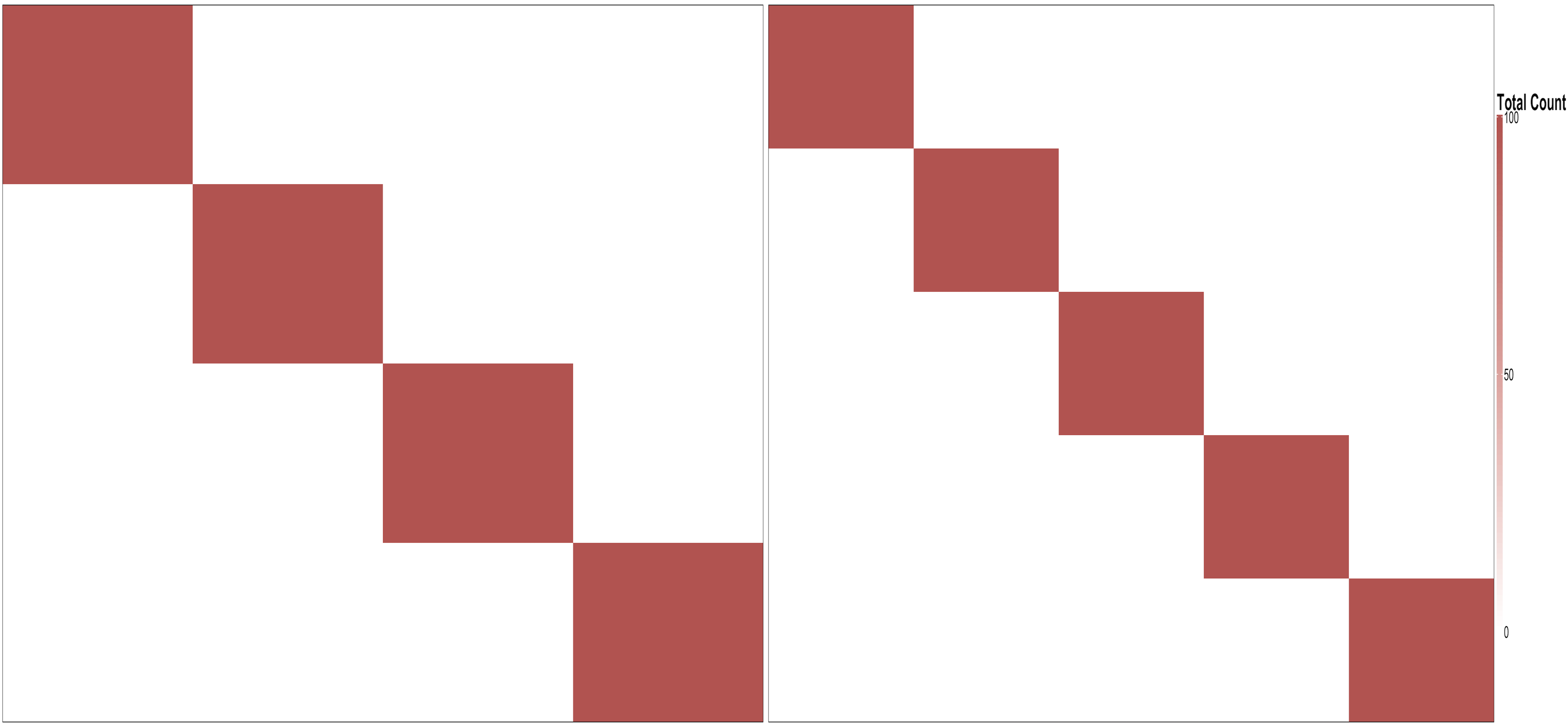}
    \caption{Covaraince matrices from generated data and the heatmap of the count for a non-zero entry within the covariance matrix for simulations of $N = 500$ given $d = 20$ with $G = 2$ and $K = 4$; $5$ respectively.}
    \label{fig:Vary d20 cov matrices}
\end{figure}

Similar to results found in Simulation Setting 1 - the algorithm was able to correctly recover the covariance matrix structure 100\% of the time with an average ARI of 1.00 for the row clusters. Heatmaps of the covariance recovery matrices are shown in Figure~\ref{fig:Vary d10 cov matrices} and Figure \ref{fig:Vary d20 cov matrices}. Perfect recovery of the covariance structure is observed despite the groups having differing column cluster configuration. With computational efficiency still comparable to the previous algorithm in Simulation Setting 1, this variation of the algorithm has additional flexibility for allowing column-groups $K_g$ to vary amongst the groups which can have substantial impact on the application.

\FloatBarrier


Overall, both PG-LBM and BMM are unable to successfully capture the data's true covariance structure for the varying $K$ case presented in simulation setting 2. Both approaches failed to recover the underlying row cluster structure and the column group structure in simulation study 11. On the other hand, the PG-LBM is able to recover the correct row cluster membership with an average ARI of 0.99 for simulation study 12 whereas BMM selected a model with a higher number of components with an average ARI of 0.48. 
Note that the BMM also allows the number of column groups to vary between the row-clusters whereas the current PG-LBM algorithm utilzied in \texttt{cobiclust()} is not currently developed to consider different column-group structures across row-clusters; hence the algorithm's inability to correctly identify and recover the true covariance structure in this setting. Interestingly enough, despite BMM identifying the correct column group structures in some groups, it has a higher misclassification rate for column grouping compared to PG-LBM. 

\FloatBarrier
\section{Breast Invasive Carcinoma Analysis}

The proposed biclustering method was then applied to RNA-seq data from breast invasive carcinoma from “The cancer genome atlas pan-cancer analysis project" (TCGA)\cite{Cancer_Genome_Atlas_Research_Network2013-an}. Two variations of the real data analysis were conducted in order to demonstrate how fixing or varying the $K$ values impacts the algorithms effectiveness. Normalization factor was computed using the trimmed mean of M values \texttt{(TMM)} approach implemented in the R package \texttt{edgeR} and incorporated in the model. In both analyses, we used the top 75 genes selected with the largest interquartile range, on the log-scale. 

\textit{Breast Invasive Carcinoma Analysis 1.} Based on the dimensionality of the dataset, we fit the model with $G=1,\ldots,8$ and $K=1,\ldots,35.$ Each group, $G$, have the same number of $K$ selected; therefore, replicating the design of simulation setting 1. 

\begin{figure}[!htb] 
\centering
    \includegraphics[width=0.9\textwidth]{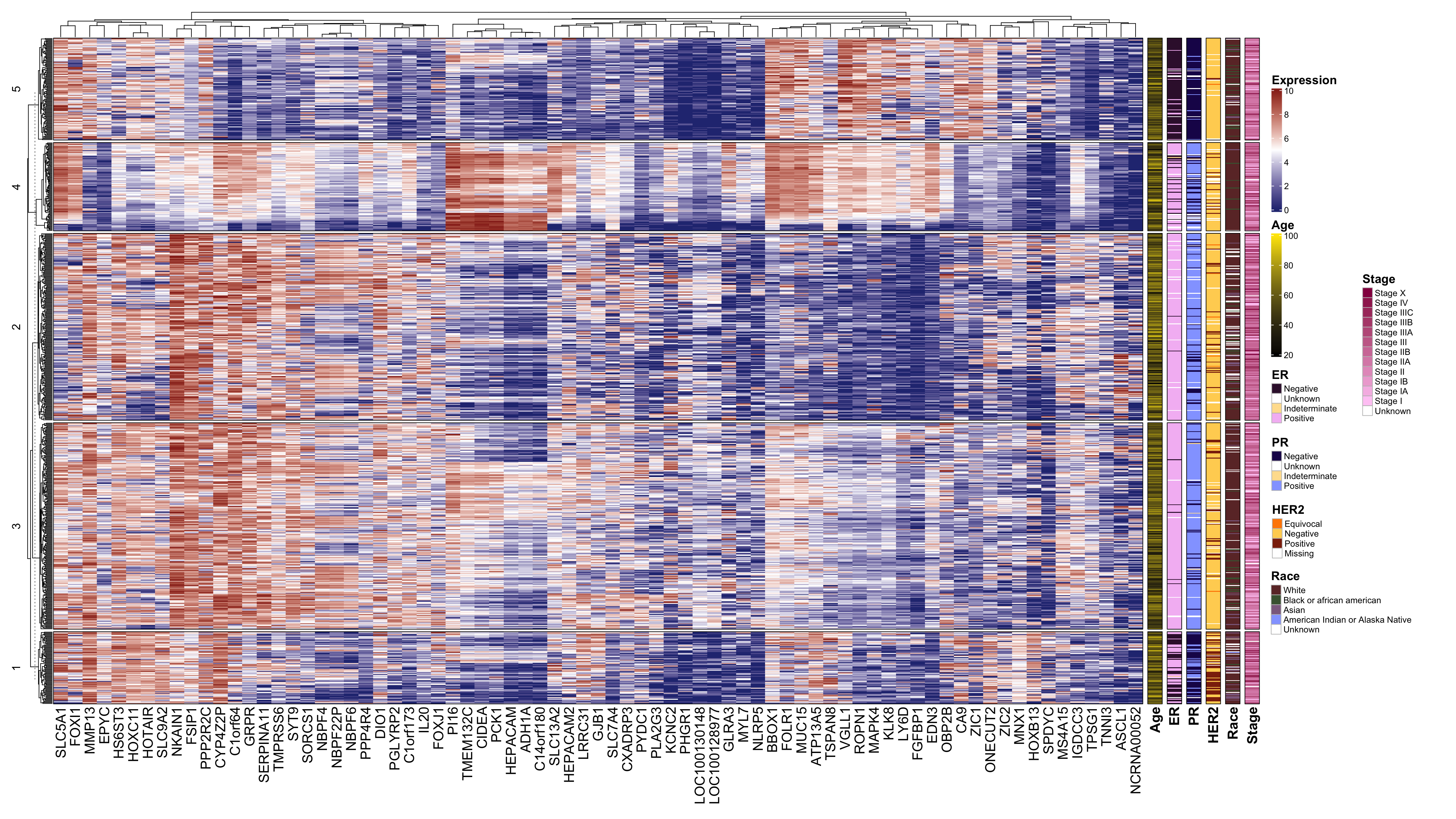}
    \caption{Heatmap showing, for Cluster 1 (C1) through Cluster 5 (C5), log-transformed gene expression patterns. This is for the $G = 5$, $K = 26$ model selected by BIC for a general $K$ value across all groups for the breast cancer invasive carcinoma RNA-seq dataset ($n = 772$). The red and blue colours represent the expression levels, such that red represents high expression and blue represents low expression. The columns and rows represent the top 75 most variable genes included in the study. ER is estrogen receptor, PR is progesterone receptor, and HER2 is human epidermal growth factor receptor 2.}
    \label{fig: General clin}
\end{figure}

\begin{figure}[!htb] 
\centering
    \includegraphics[width=0.9\textwidth]{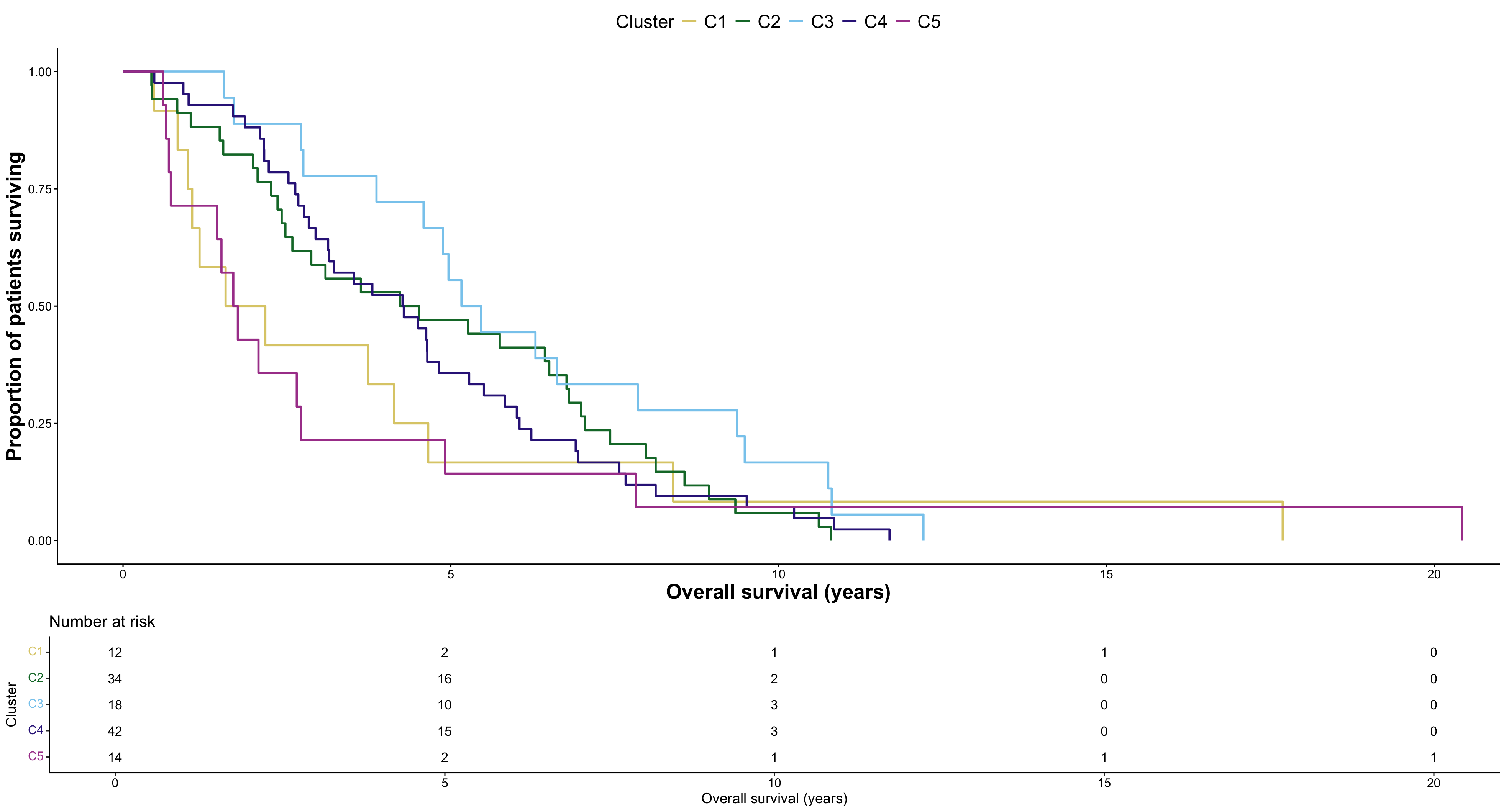}
    \caption{Kaplan-Meier plot of overall survival for Cluster 1 (C1) through Cluster 5 (C5) for the $G = 5$, $K = 26$ model selected by BIC for a general $K$ value for the breast cancer invasive carcinoma RNA-seq dataset.}
    \label{fig:General surv plot}
\end{figure}

\begin{figure}
      \centering
	   \begin{subfigure}{0.45\linewidth}
		\includegraphics[width=\linewidth]{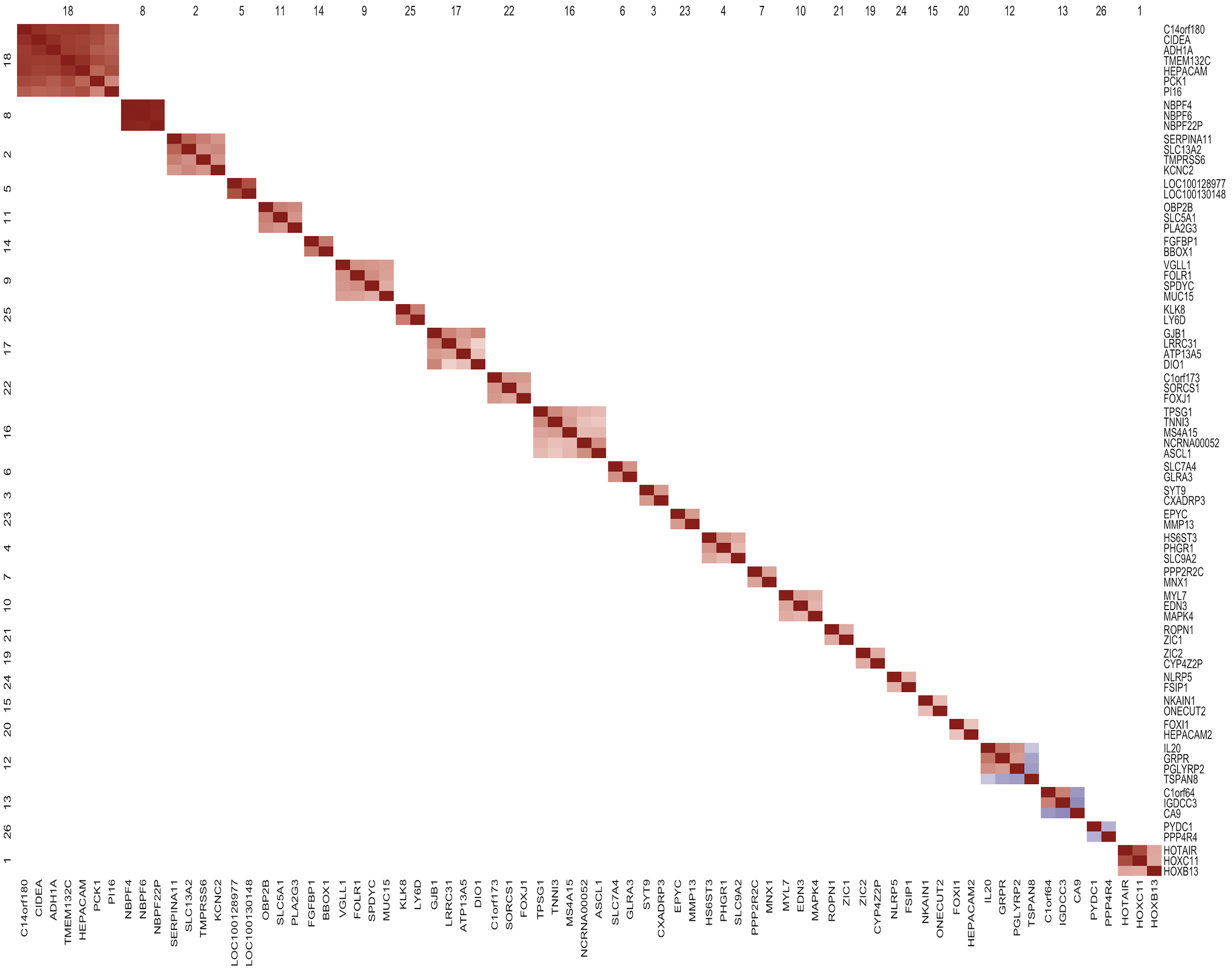}
		\caption{Recovered correlation matrix 1}
		\label{fig:gen_cov1}
	    \end{subfigure}
	     \begin{subfigure}{0.45\linewidth}
		 \includegraphics[width=\linewidth]{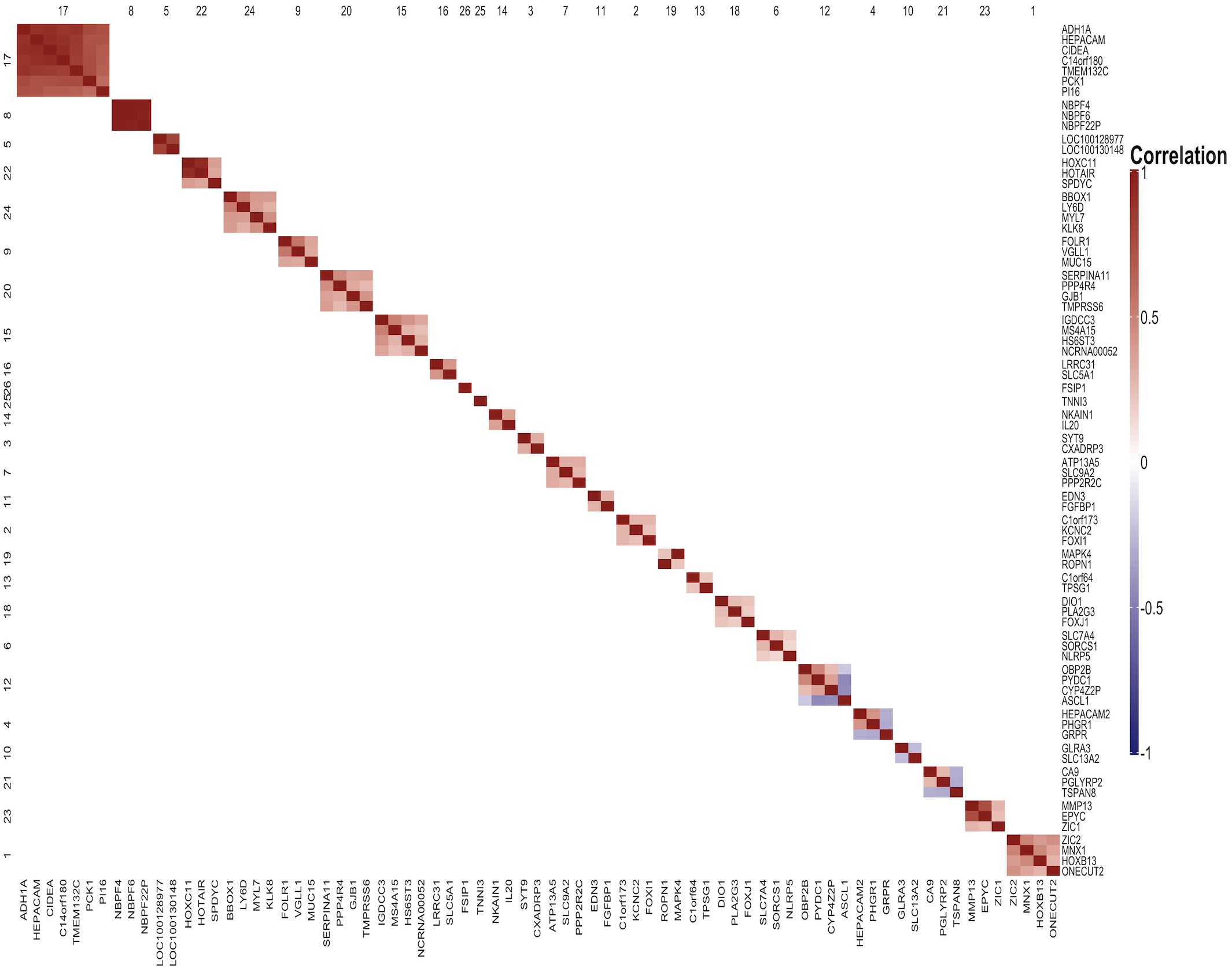}
		 \caption{Recovered correlation matrix 2}
		 \label{fig:gen_cov2}
	      \end{subfigure}   
       \vfill
	   \begin{subfigure}{0.45\linewidth}
		  \includegraphics[width=\linewidth]{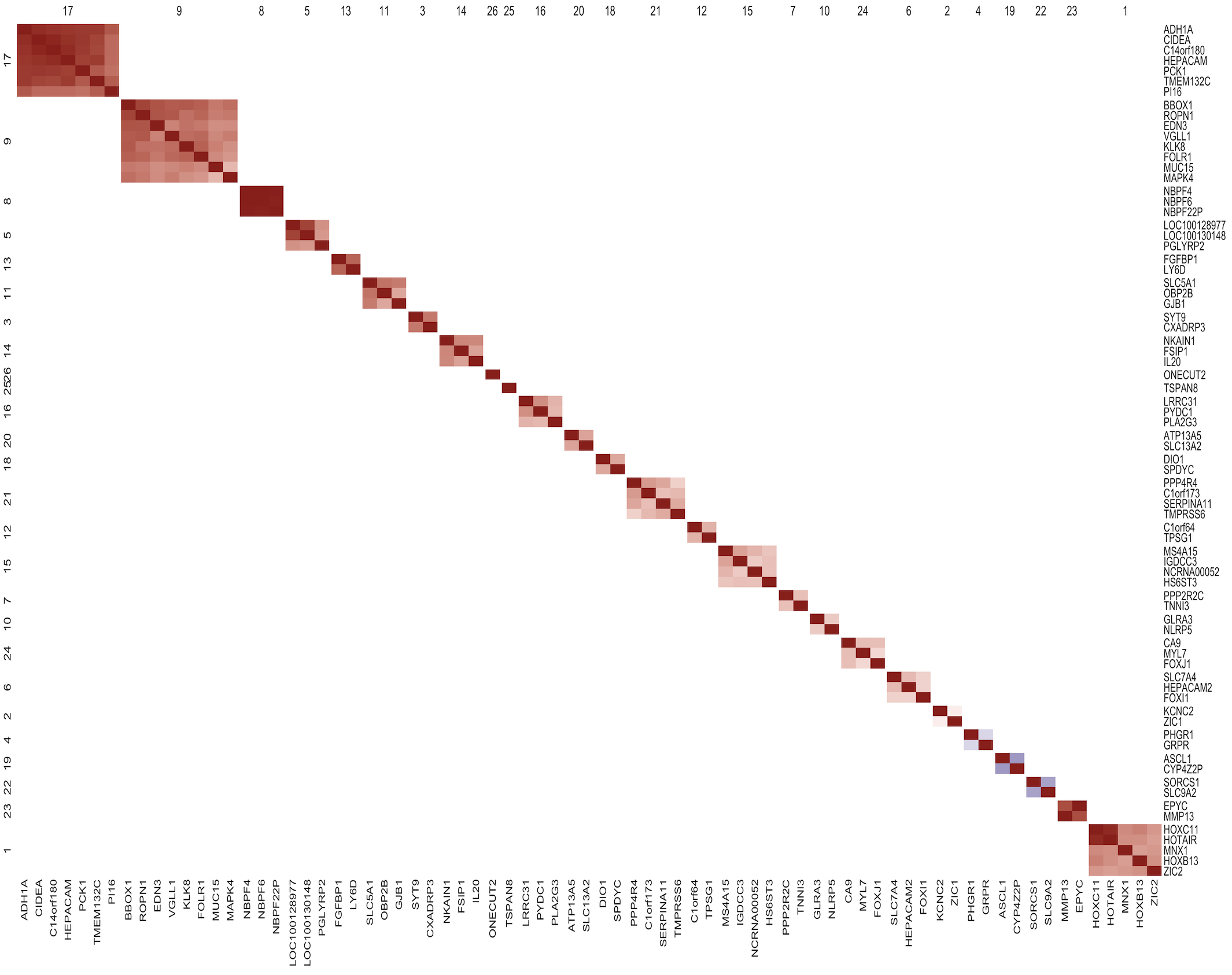}
		  \caption{Recovered correlation matrix 3}
		  \label{fig:gen_cov3}
	       \end{subfigure}
	     \begin{subfigure}{0.45\linewidth}
		 \includegraphics[width=\linewidth]{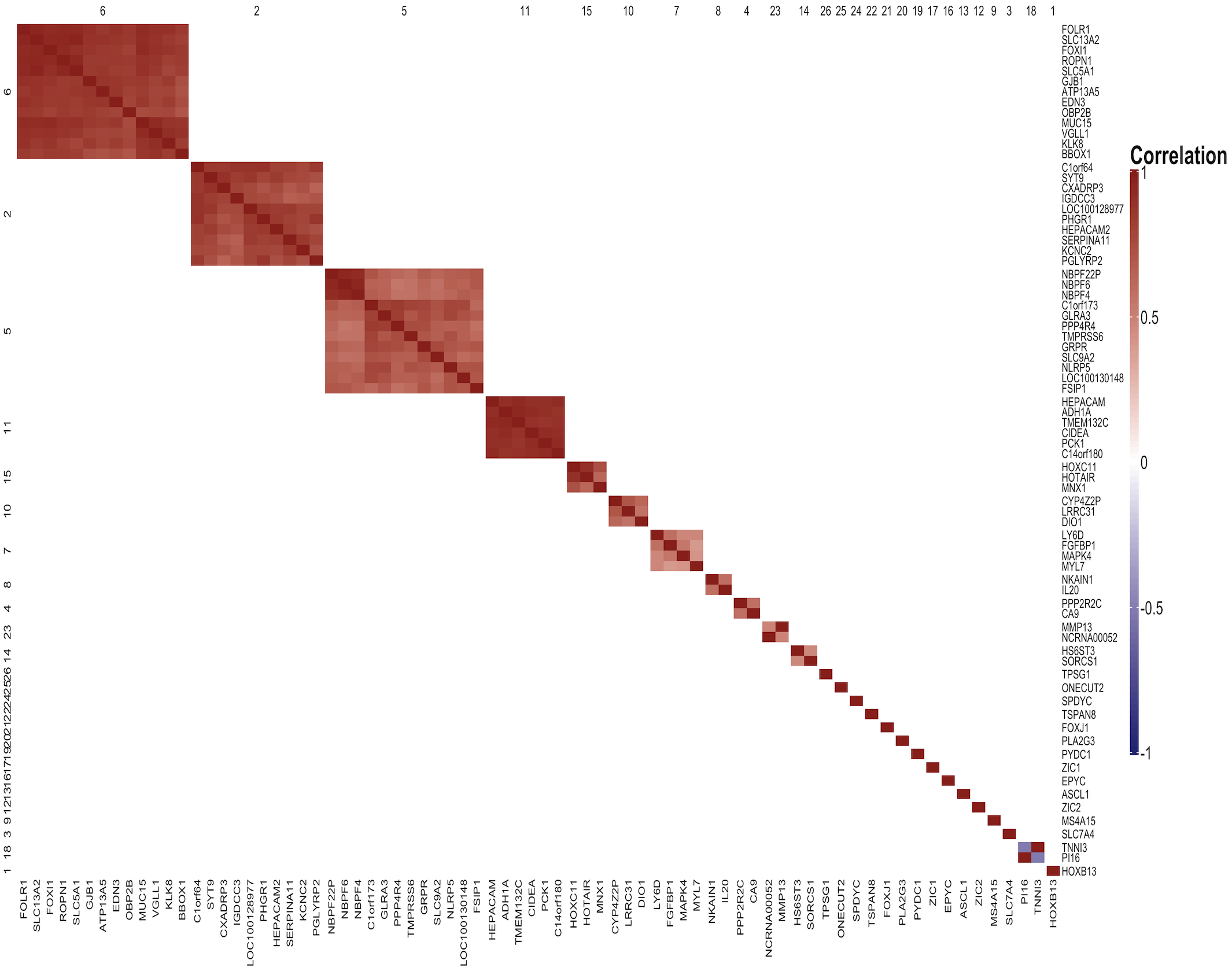}
		 \caption{Recovered correlation matrix 4}
		 \label{fig:gen_cov4}
	      \end{subfigure}
       \vfill
	       \begin{subfigure}{0.45\linewidth}
		  \includegraphics[width=\linewidth]{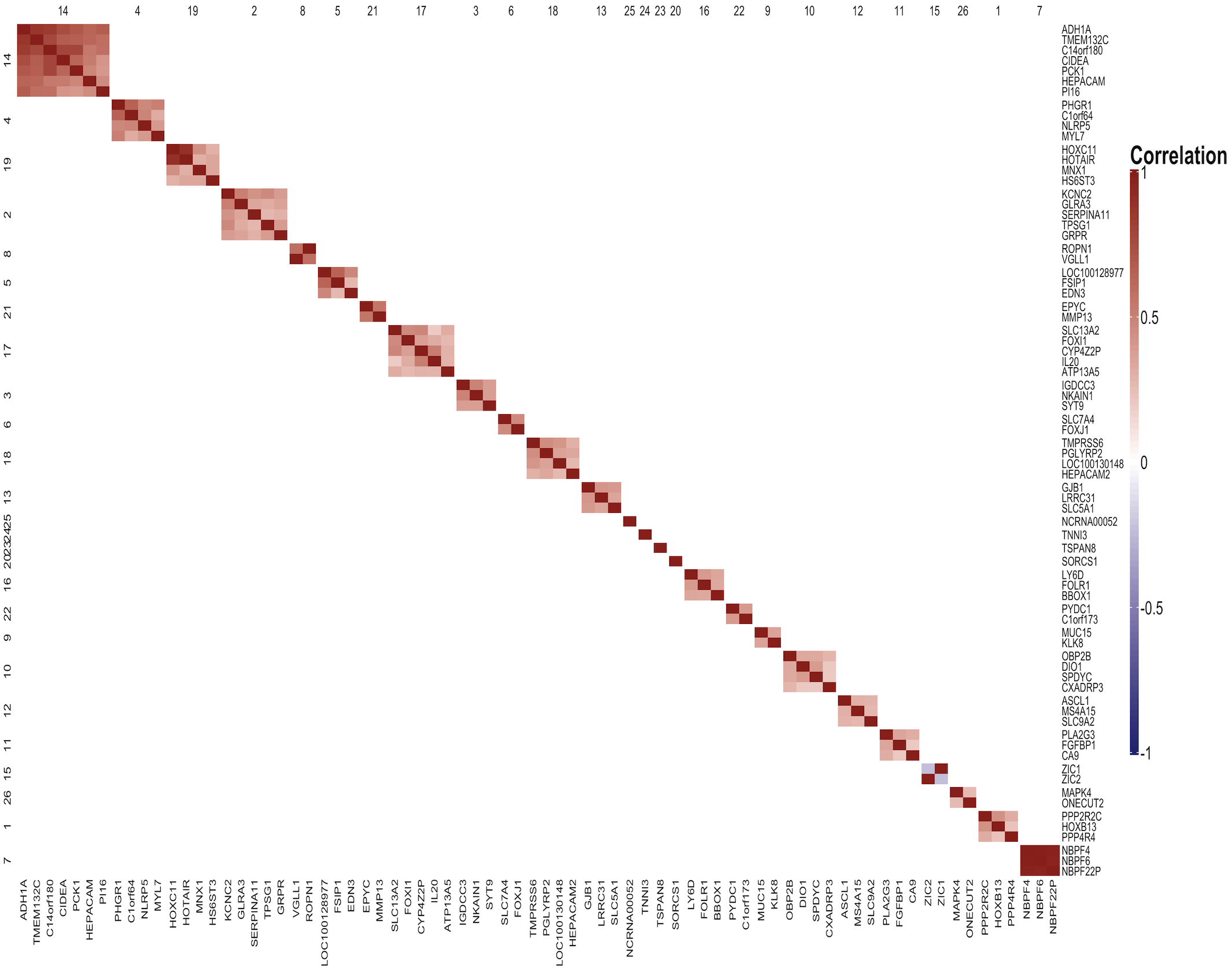}
		  \caption{Recovered correlation matrix 5}
		  \label{fig:gen_cov5}
	       \end{subfigure}
	\caption{Heatmap showing, for Cluster 1 (C1) through Cluster 5 (C5), recovered covariance matrices for gene expression patterns. This is for the $G = 5$, $K = 26$ model selected by BIC for a general $K$ value across all groups for the breast cancer invasive carcinoma RNA-seq dataset (n = 772). The red and blue colours represent the correlation levels, where red represents high gene correlation and blue represents low gene correlation with column clusters visualized on the diagonal of the recovered covariance matrix.}
\label{fig: Gen_Results}
\end{figure}

The results display that a $G = 5$ model with $K = 26$ was selected as the best-fitting model. Heatmaps of the gene-expression patterns and covariance recovery matrices are shown in Figure~\ref{fig: General clin} and Figure~\ref{fig: Gen_Results}. The overall survival (OS) of the five groups is shown in Figure~\ref{fig:General surv plot}. Cluster 1 and Cluster 5 had similar OS trends, with many observations surviving shorter timeframes than Clusters 2-4. However, both Clusters 1 and 5 had a single observation survive longer than 15 years which pulled their OS ahead of the remaining groups. Clusters 2, 3 and 4 had a similar survival trend with Cluster 2 having the worst OS overall. Interestingly enough, both Cluster 1 and Cluster 5 display a high proportion of observations with negative ER and PR expression compared to Clusters 2-4 which mainly exhibited positive ER and PR expression. Literature has stated that both positive ER and PR statuses have shown to be correlated with favourable patient outcomes \cite{Davey2021}. This is generally seen as the OS has a more immediate decrease in Clusters 1 and 5 which contain a higher presence of negative ER and PR status compared to the remaining clusters which presented higher proportions of positive ER and PR status. Since PR is an estrogen-regulated gene, ER-positive tumors are typically PR-positive and ER-negative tumors are PR-negative \cite{Bae2015}. This trend is evident in Cluster 2, 3 and 4 where a majority of cases have both positive ER and PR statuses. Group 1 is the only cohort to demonstrate a trend of positive HER2 expression with all other groups showing high quantities of negative HER2 expression. In breast cancer, HER2 overexpression or HER2 positive status has corresponded with poor prognosis and decreased survival \cite{CHEN2022}. When examining racial trends, Cluster 1 appears to contain a higher population of individuals identifying as Asian whereas Clusters 2-5 do not present any visual racial trends. The distribution of age and stage of disease appears to be even, with no specific group containing a high quantity of individuals with a particular age or specific disease stage. 

Expression patterns amongst the top 75 genes help to illustrate expression differences between groups and may highlight potential genes of interest that improve or reduce the OS of observations. The expression patterns found here were supported in the literature. We selected some genes to illustrate this. For example, NKAIN1 is a development-related gene that shows high expression across all clusters. In humans, NKAIN1 encodes the sodium/potassium transporting ATPase interacting with 1 protein. Literature has found an association between high expression of NKAIN1 and breast cancer and NKAIN1 mRNA levels have been recognized as a possible biomarker in the diagnosis of breast cancer \cite{Ramezani2022}. Furthermore, HS6ST3 is the Heparan Sulfate 6-O-Sulfotransferase 3 protein-coding gene. Evidence has shown that it is highly expressed in breast cancer cell lines and through its silencing tumor growth and progression were diminished \cite{Iravani2017}. Interestingly, Clusters 1, 4 and 5 showed high expression of HS6ST3, Cluster 3 showed intermediate expression and Cluster 2 showed the lowest expression. This may explain why Cluster 1 and 5 had the quickest drop in OS despite having observations survive past the 15-year time-stamp. In addition, the metabolism-related gene, C1orf64 encodes the steroid receptor-associated and regulated protein. Findings have indicated that the C1orf64 gene has high expression in breast tumours \cite{Naderi2017}. Here, Cluster 1 has moderate expression of C1orf64 while Cluster 5 has low expression which is in accordance with OS observed. Neurological-related genes have also shown a correlation with breast cancer diagnosis and prognosis. For instance, ZIC1 is a zinc finger of the cerebellum 1 gene and is known to have a tumour-suppressive role in breast cancer \cite{Han2018}. Evidence shows that high expression of the ZIC1 gene suppresses the growth of breast cancer cells and xenograft tumours \cite{Han2018}. Interestingly, most clusters display low expression whereas Cluster 5 showed high expression, supporting the OS differences observed.


The recovered covariance matrices generally demonstrate moderate to strongly correlated clusters of the selected genes across all groups; however, many of the column-groups were quite small in size. A majority of the column clusters contained two to four genes indicating that a larger $K$ value was needed. By restricting the row-groups to have the same number of column-groups, the algorithm may be imposing that the model inflates the number of column-groups necessary in order to maximize the model selection criteria. The selection of a larger $K$ value indicates that perhaps more clusters containing highly correlated genes have been generated - and thus more specified clusters have been produced. 

\textit{Breast Invasive Carcinoma Analysis 2.} By assuming that the size of $K$ remains the same across all groups, a restrictive covariance structure is being forced onto the data. By mirroring simulation setting 2, we allow the $K$ values to vary across the different groups and we fit the model with $G=1,\ldots,8$ and a maximum column group value of $K=15$.

\begin{figure}[!htb] 
\centering
    \includegraphics[width=0.9\textwidth]{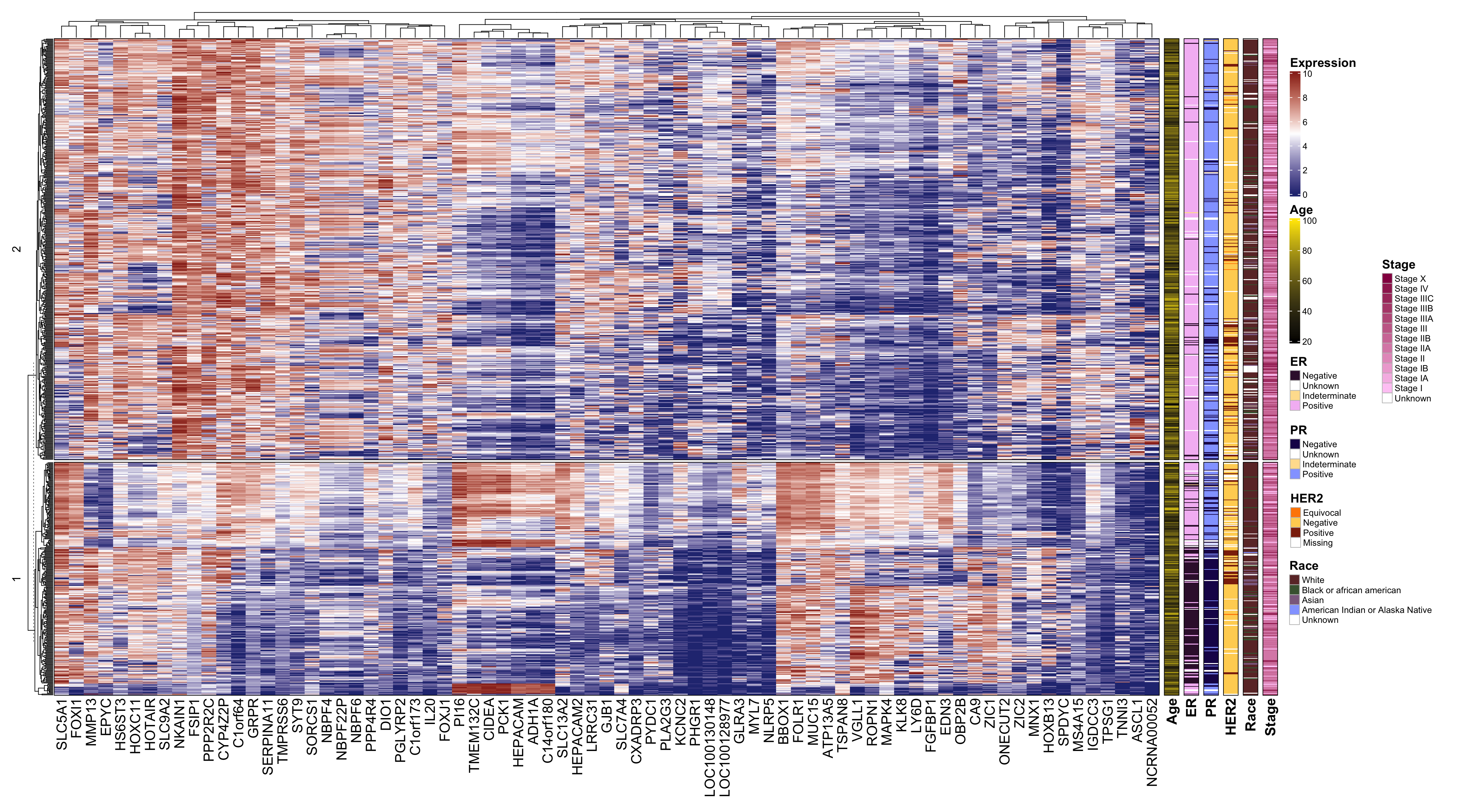}
    \caption{Heatmap showing, for Cluster 1 (C1) through Cluster 2 (C2), log-transformed gene expression patterns. This is for the $G = 2$, $K = 5$ and $K = 4$ model selected by BIC for a varying $K$ value across all groups for the breast cancer invasive carcinoma RNA-seq dataset (n = 772). The red and blue colours represent the expression levels, where red represents high expression and blue represents low expression. The columns and rows represent the top 75 most variable genes included in the study. ER is estrogen receptor, PR is progesterone receptor, and HER2 is human epidermal growth factor receptor 2.}
    \label{fig: Vary clin}
\end{figure}

\begin{figure}[!htb] 
\centering
    \includegraphics[width=0.9\textwidth]{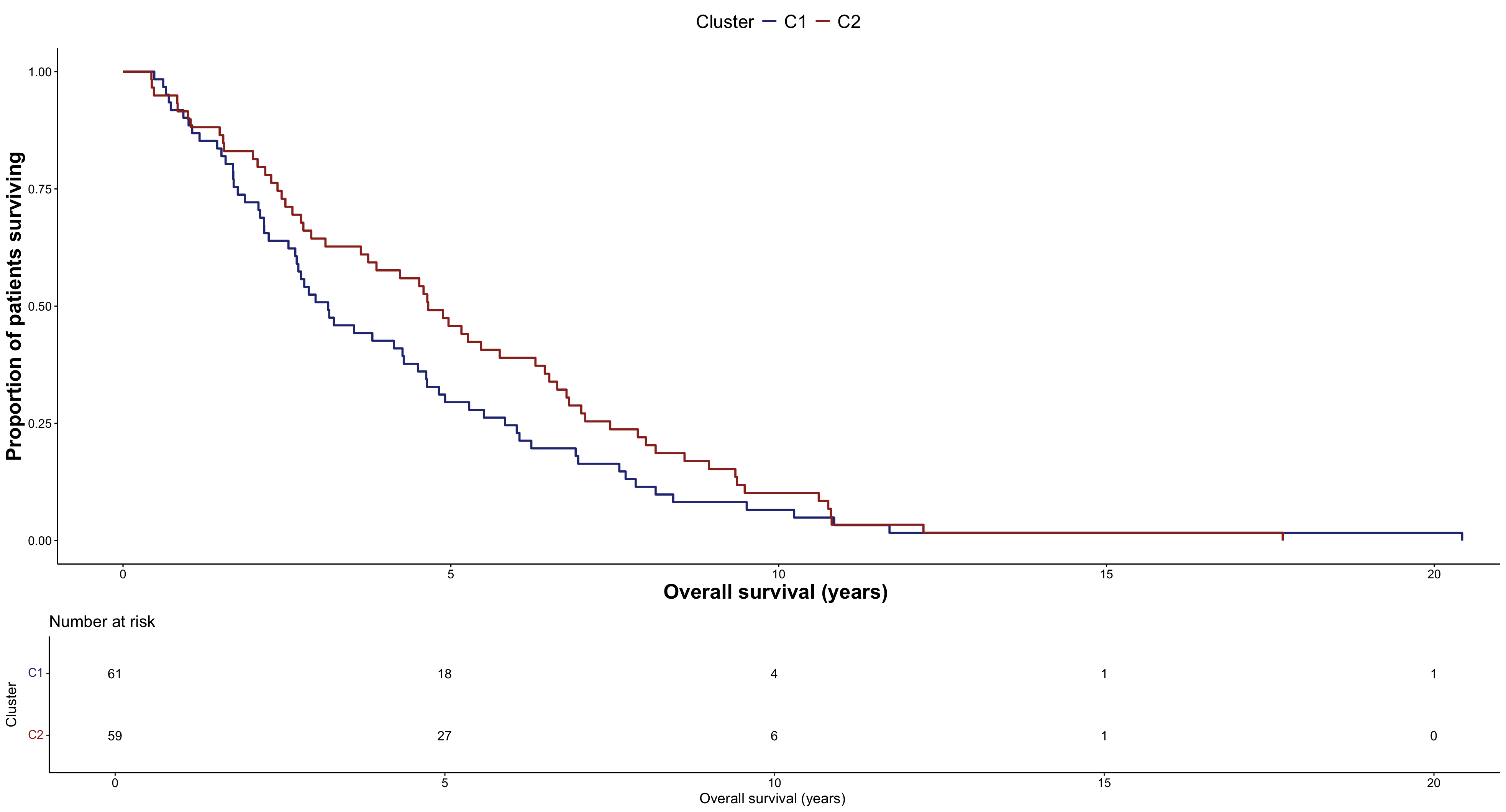}
    \caption{Kaplan-Meier plot of overall survival for Cluster 1 (C1) through Cluster 5 (C5) for the $G = 2$, $K = 5,4$ model selected by BIC for a varying $K$ value for the breast cancer invasive carcinoma RNA-seq dataset.}
    \label{fig:Vary surv plot}
\end{figure}

\begin{figure}
      \centering
	   \begin{subfigure}{0.48\linewidth}
		\includegraphics[width=\linewidth]{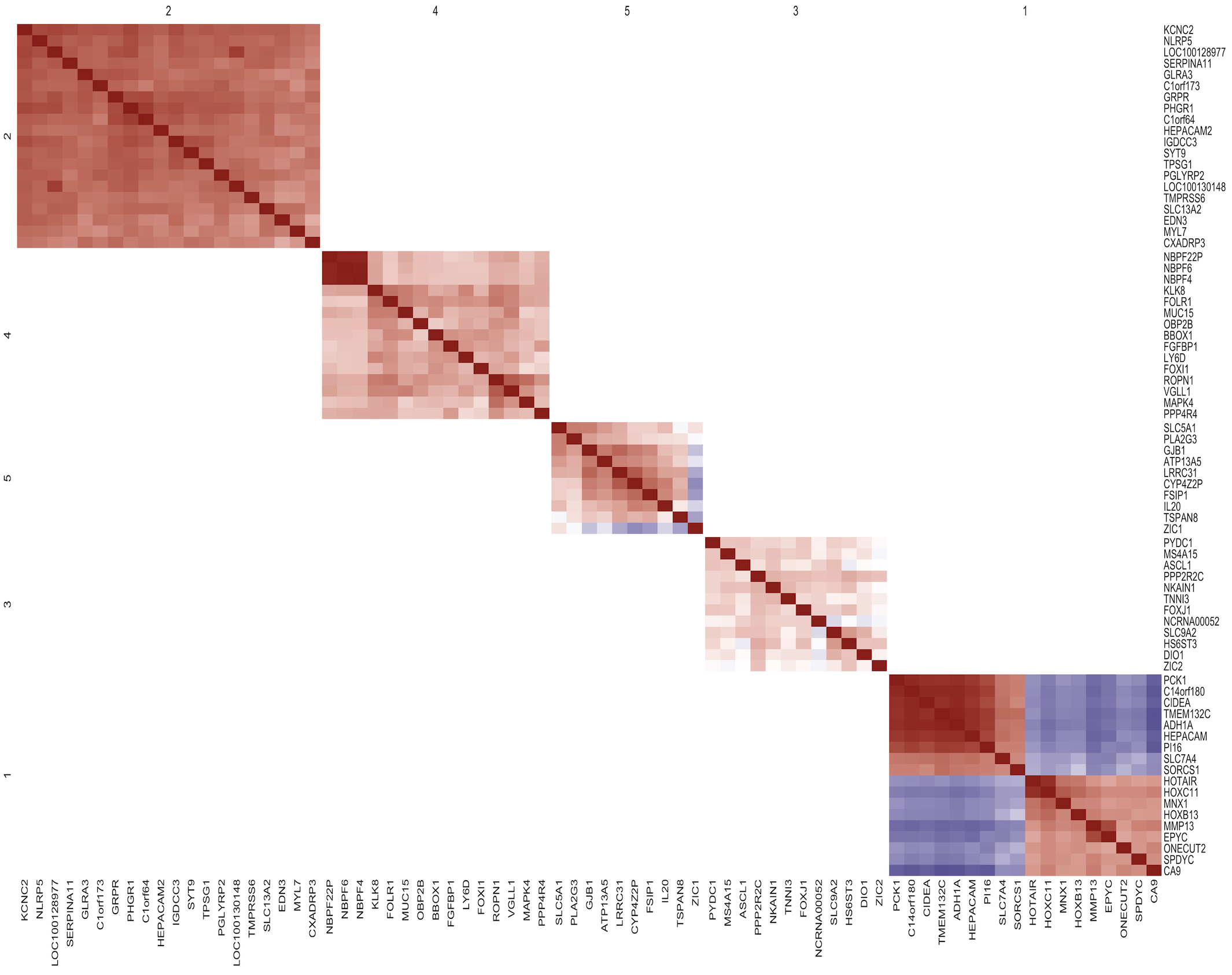}
		\caption{Recovered correlation matrix 1}
		\label{fig:vary_cov1}
	    \end{subfigure}
	     \begin{subfigure}{0.48\linewidth}
		 \includegraphics[width=\linewidth]{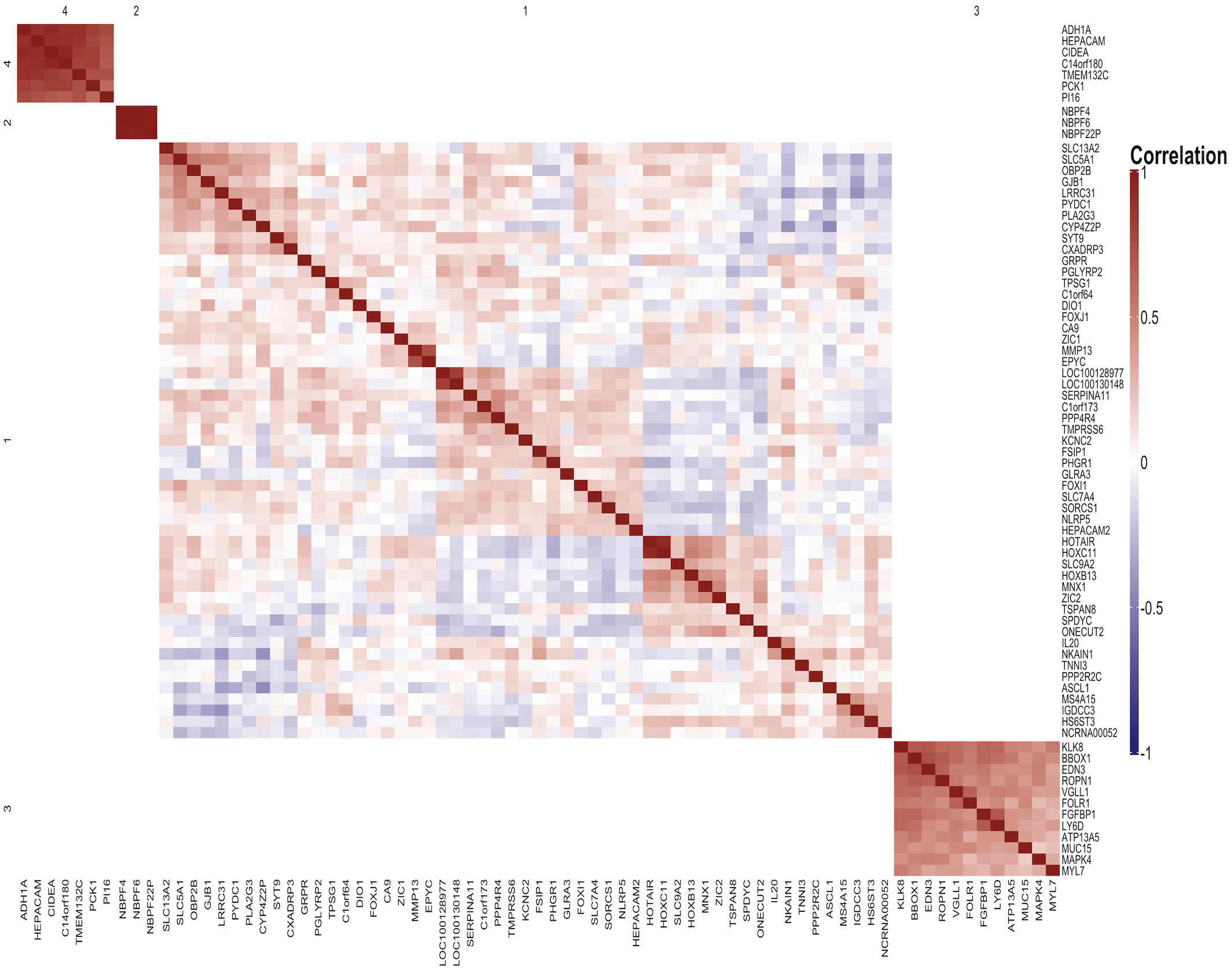}
		 \caption{Recovered correlation matrix 2}
		 \label{fig:vary_cov2}
	      \end{subfigure}
	\caption{Heatmap showing, for Cluster 1 (C1) through Cluster 2 (C2), recovered covariance matrices for gene expression patterns. This is for the $G = 5$, $K = 5$ and $K = 4$ model selected by BIC for a general $K$ value across all groups for the breast cancer invasive carcinoma RNA-seq dataset (n = 772). The red and blue colours represent the correlation levels, where red represents high gene correlation and blue represents low gene correlation with column clusteres visualized on the diagonal of the recovered covariance matrix.}
	\label{fig: Vary_Results}
\end{figure}

The results display that the $G = 2$ model with $K_1$ = 5 and $K_2$ = 4 was selected as the best-fitting model. Heatmaps of the gene-expression patterns and covariance recovery matrices are shown in Figure~\ref{fig: Vary clin} and Figure~\ref{fig: Vary_Results}. Similar to Breast Invasive Carcinoma Analysis 1, Cluster 1 displays a high proportion of observations with negative ER and PR expression compared to Cluster 2 which mainly exhibits positive ER and PR expression. Furthermore, Cluster 2 exhibits the trend of having both a positive ER and PR status for a majority of the cases. However, unlike Breast Invasive Carcinoma Analysis 1, HER2 expression does not appear to differ across the two groups. Similar to Cluster 1 and Cluster 5 in Breast Invasive Carcinoma Analysis 1, Cluster 1 has the longest OS time but the general OS decreases faster than Cluster 2 which could be attributed in part to the high presence of cases with a negative ER and PR status. Exploring the presence of racial trends displays that Cluster 1 has a higher population of individuals identifying as Asian. The age and stage of disease trends did not change between analyses, where there continued to be no specific group containing a high quantity of individuals with a particular age or specific disease stage. The OS of the two groups are shown in Figure~\ref{fig:Vary surv plot}. Both groups had a similar OS trend, however, Cluster 2 observed an unfavourable OS despite having a higher OS for the first 12 years. 

The expression patterns amongst the top 75 genes can also be examined for this analysis. For NKAIN1, HS6ST3, and C1orf64, Cluster 2 displayed higher expression levels for all three of these genes compared to Cluster 1 which supports the difference in OS observed. Furthermore, ZIC1 shows moderate expression in Cluster 1 and low expression in Cluster 2, agreeing with the OS observed in the current study.   

Unlike the previous analysis, the column-groups were allowed to vary across the groups which lent a more flexible framework. The flexibility allowed for both a smaller row-cluster and column-group value to be selected while still maximizing the selection criterion. Negatively correlated genes were easier to identify in this analysis for both groups. Group 1 demonstrates a strong negative correlation amongst nine different genes which was not observed in Breast Invasive Carcinoma Analysis 1. The results demonstrate that by allowing the column-groups to vary across the row-clusters, the algorithm selected a more generalized cluster structure as opposed to the more specified structure as seen in the previous analysis. 

\FloatBarrier
\section{Discussion}

A model-based bi-clustering approach utilizing a mixture of MPLN distribution and a block-diagonal covariance structure is introduced. To our knowledge, this is the first use of a block-diagonal covariance structure in the literature. The implementation of the proposed biclustering algorithm in both the equal $K$ and varying $K$ across groups simulation settings showed strong performance accuracy and efficiency. In both settings, results demonstrated accurate row clustering with high ARI values as well as precise recovery of the covariance structure. Furthermore, BIC selected the final models with the true $G$ and $K$ values of the simulated data. The algorithm performs well when there is a large number of observations as well as high dimensionality. However, authors' note that this algorithm is not designed for the scenario when the number of features is significantly higher than the number of observations. The proposed algorithm in comparison to PG-LBM and BMM highlighted the novelty and resourcefulness of this current approach. The PG-LBM appears to be the most rigid in terms of selecting the covariance structure of the data as it is unable to select row-cluster specific column groupings, unlike the BMM and proposed frameworks. Additionally, although the BMM was more flexible in its column-group selection, its accuracy was hindered by the nature of the data being over-dispersed whereas the PG-LBM and MPLN are designed to handle this data type. Furthermore, a major difference between the PG-LBM/BMM and the proposed method is that PG-LBM and BMM only permit positive correlations within the column clusters. The use of an MPLN framework allows negative correlations since the covariance between two components of a random variable follows a random variable with a multivariate normal distribution. In particular, this is an important component when considering the algorithm's applications in a bioinformatics setting. It may be of interest to group variables based on the magnitude of correlation regardless of the sign of correlation. For instance, a certain pathway may have a critical role in tumor development and thus, the genes involved in that pathway show expression level changes. Some genes may be up-regulated while other genes are down-regulated, and therefore, the genes will be divided into multiple column clusters which could lead to an overestimation of the number of column clusters. Moreover, the column clusters will only provide a partial and deficient outline of the pathway's involvement. 

Overall, applying both variants of the algorithm identified a differing number of subtypes of breast invasive carcinoma that vary from each other in terms of patient survival outcomes, clinical covariates and gene expression. In the case with equal column-groups across groups, the algorithm identified 5 subtypes of breast invasive carcinoma whereas the varying column-groups identified 2 subtypes. By restricting the number of column-groups to be the same amongst the groups, the algorithm requires a large maximum column-group value to select the best-fitting model. To maximize BIC, a larger and smaller number of row-clusters and column-groups, respectively, are selected to ensure consistent correlation trends across all the groups. When the column-group value is allowed to vary across the groups, the BIC selected a more generalized structure with both fewer row- and column-groups. The flexibility of having the column-groups vary across row-clusters led to larger column-group sizes with the smallest column-group now containing 10 genes. Associations between gene expression patterns, clinical covariates and patient outcomes are important when understanding disease diversity. The findings in this study highlight the molecular and clinical heterogeneity of breast cancer which may help inform the development of subtype-specific treatment procedures to further improve patient outcomes. 

The proposed approach assumes that a variable can only belong to one of the column-groups. This may be too restrictive when dealing with RNA-seq data. A gene could be involved in more than one pathway. Future research will focus on extending the proposed approach to allow for soft column-group assignments as opposed to hard column-group assignments.


\section{Acknowledgements}
This work was supported by the NSERC Discovery Grant (2021-03812 (Subedi); 2018-04444 (Browne)); and funding from the Canada Research Chairs Program (2020-00303 (Subedi)). This research was enabled in part by support provided by Research Computing Services (https://carleton.ca/rcs) at Carleton University.

\newpage
\begin{appendices}

\FloatBarrier
\section{Details on Parameter Recovery}\label{para}

\begin{center}
\begin{table}[!htb]
\centering
 \caption{Summary of parameter recovery performance of proposed biclustering framework in all ten simulation settings.} 
\scalebox{0.80}{
\begin{tabular}{@{\extracolsep{4pt}}cccccccc@{}}
\cline{1-8}
N&Model&\multicolumn{3}{l}{MSE of $\bmu_g$}&\multicolumn{3}{l}{MSE of $\piv_g$}\\
\cline{3-5}
\cline{6-8}
&&Comp 1&Comp 2&Comp 3&Comp 1&Comp 2&Comp 3\\
\cline{1-8}\\
n = 500&\begin{tabular}[c]{@{}l@{}}
d = 10 \\ K = 2
\end{tabular}&0.0084&0.0028&--&3.73e-4&3.73e-4&--\\[10pt]

&\begin{tabular}[c]{@{}l@{}}
d = 20 \\ K = 4
\end{tabular}&0.0021&0.041&--&4.85e-4&4.85e-4&--\\[10pt]

&\begin{tabular}[c]{@{}l@{}}
d = 50 \\ K = 9
\end{tabular}&0.0074&0.0025&--&3.98e-4&3.98e-4&--\\[10pt]

&\begin{tabular}[c]{@{}l@{}}
d = 50 \\ K = 10
\end{tabular}&0.0076&0.0021&--&3.54e-4&3.54e-4&--\\[10pt]

&\begin{tabular}[c]{@{}l@{}}
d = 50 \\ K = 12
\end{tabular}&0.0075&0.0023&--&3.30e-4&3.30e-4&--\\[10pt]

n = 1000&\begin{tabular}[c]{@{}l@{}}
d = 10 \\ K = 2
\end{tabular}&0.0029&0.0092&0.0015&9.26e-5&2.07e-4&2.29e-4\\[10pt]

&\begin{tabular}[c]{@{}l@{}}
d = 20 \\ K = 4
\end{tabular}&0.0030&0.0074&0.0016&1.03e-4&1.91e-4&2.06e-4\\[10pt]

&\begin{tabular}[c]{@{}l@{}}
d = 50 \\ K = 9
\end{tabular}&0.0027&0.00961&0.0014&1.10e-4&1.62e-4&2.64e-4\\[10pt]

&\begin{tabular}[c]{@{}l@{}}
d = 50 \\ K = 10
\end{tabular}&0.0027&0.0084&0.015&7.14e-5&2.18e-4&1.94e-4\\[10pt]

&\begin{tabular}[c]{@{}l@{}}
d = 50 \\ K = 12
\end{tabular}&0.0026&0.0093&0.0013&1.16e-4&2.59e-4&2.87e-4\\[10pt]
\cline{1-8}
\end{tabular}
}
\end{table}
\label{MSE result}
\end{center}

\FloatBarrier
\section{Additional Covariance Recovery Matrices}\label{vis extra}

\begin{figure}
      \centering
	   \begin{subfigure}{0.52\linewidth}
        \includegraphics[width=\linewidth]{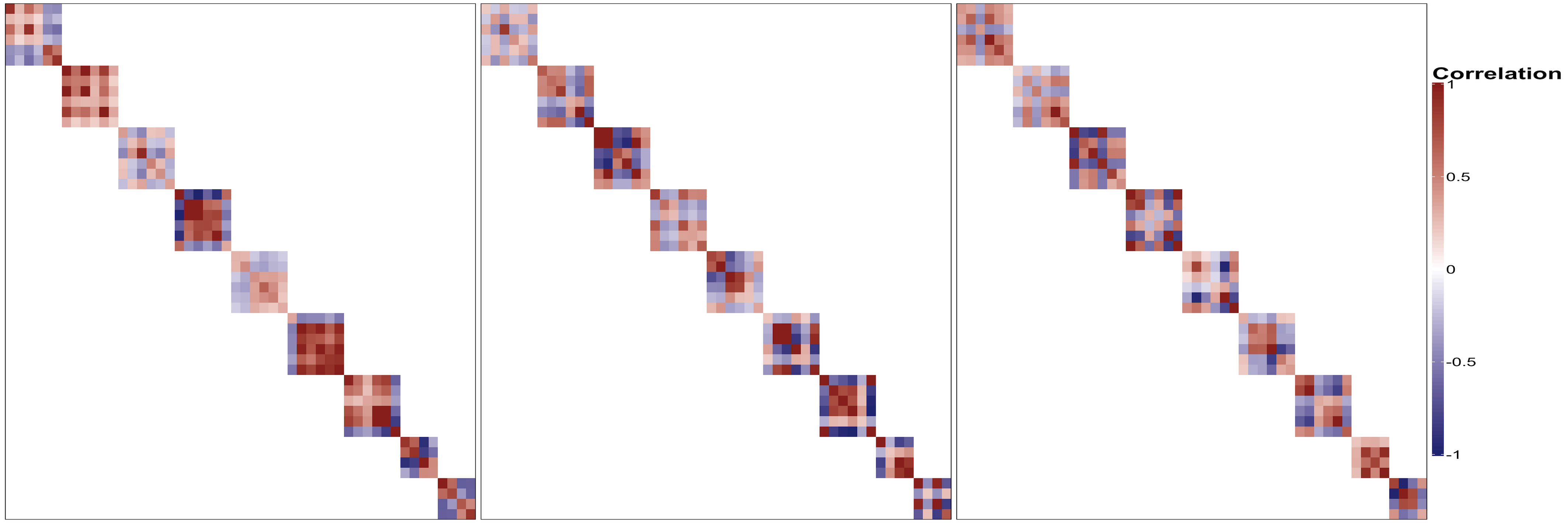}
		\includegraphics[width=\linewidth]{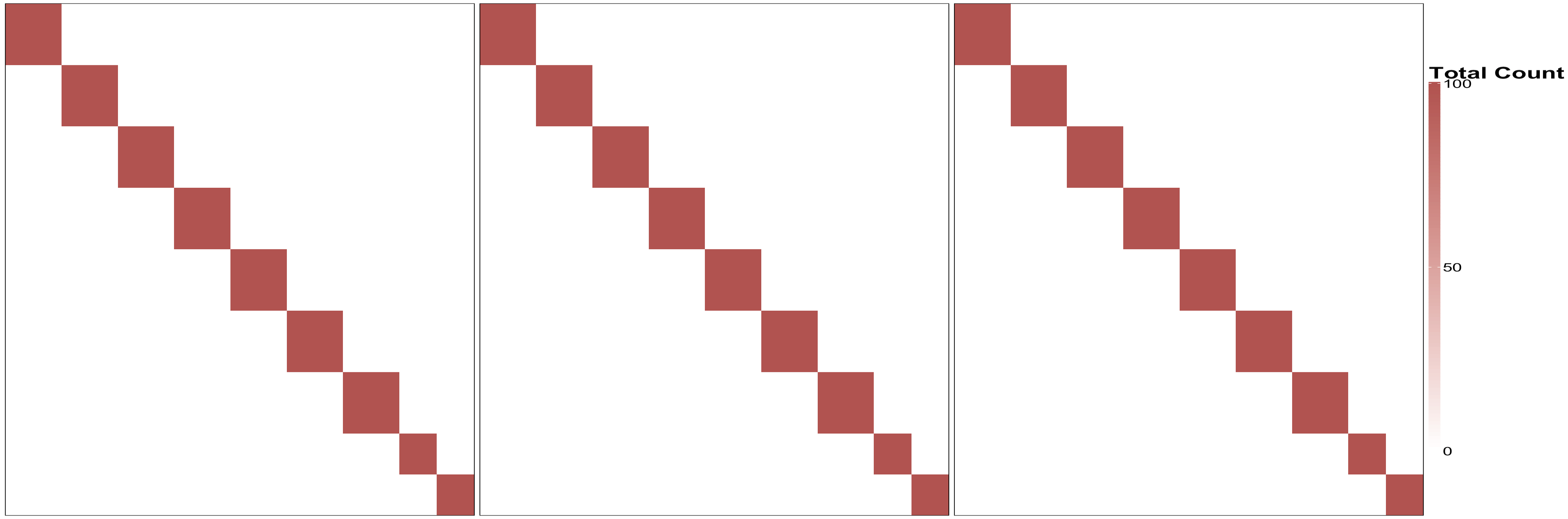}
		\caption{$N$ = 1000; $d$ = 50; $K$ = 9}
		\label{fig:d50_K9}
	    \end{subfigure}
	     \begin{subfigure}{0.52\linewidth}
         \includegraphics[width=\linewidth]{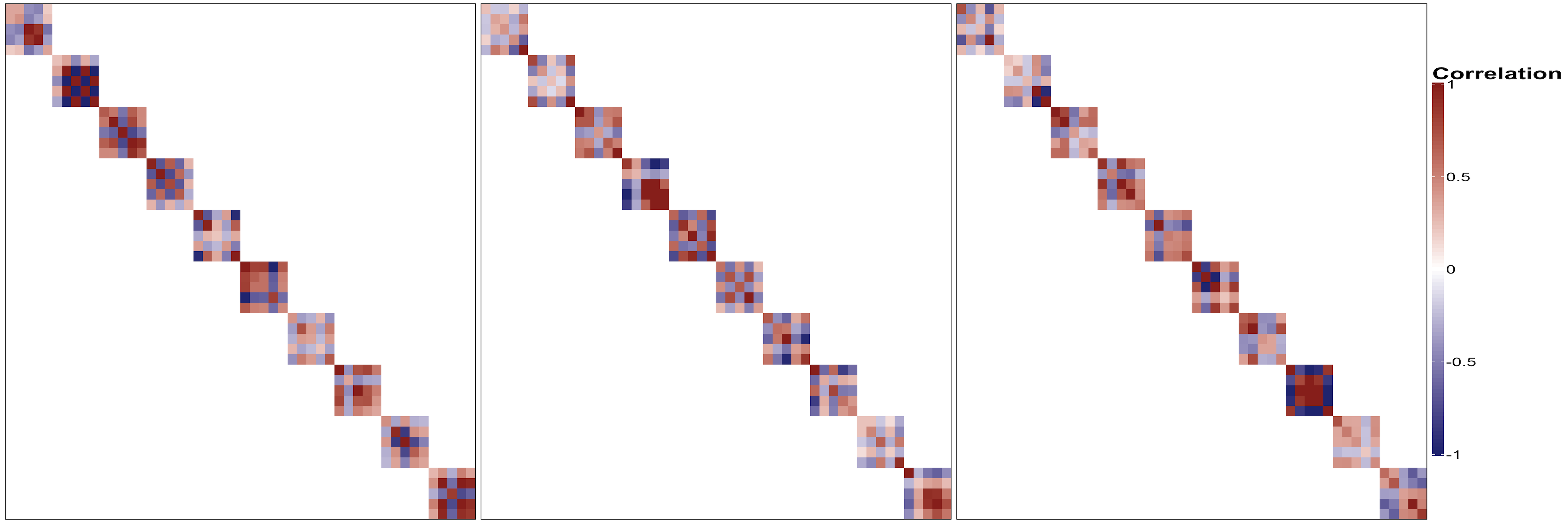}
		 \includegraphics[width=\linewidth]{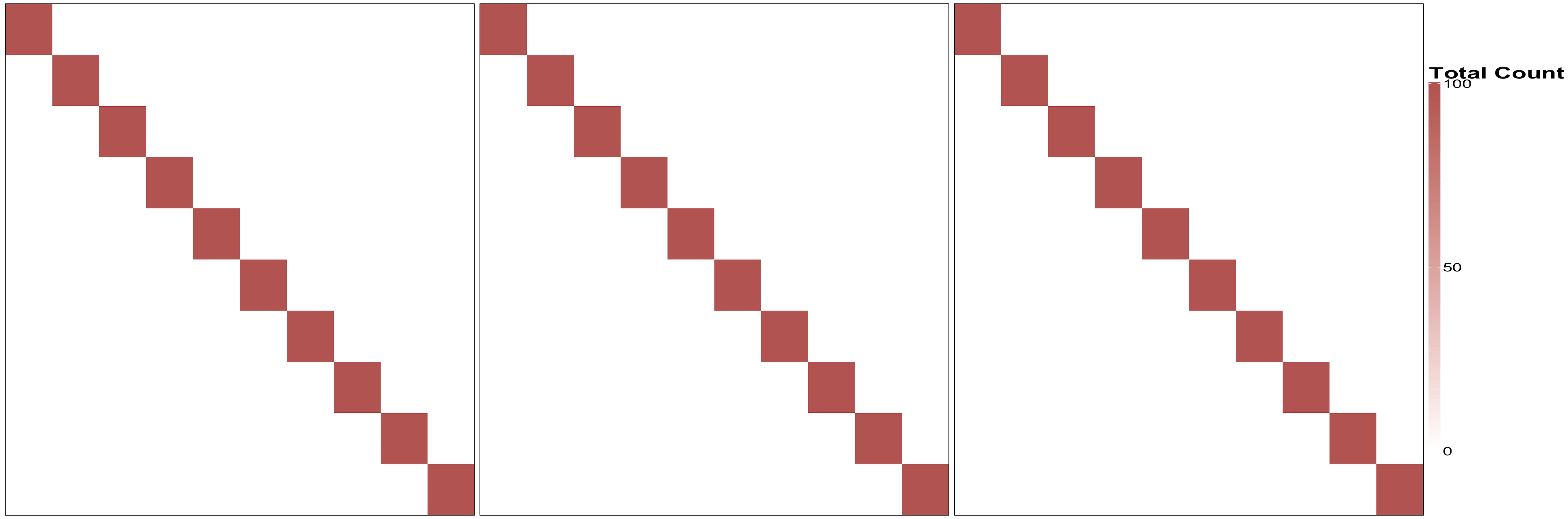}
		 \caption{$N$ = 1000; $d$ = 50; $K$ = 10}
		 \label{fig:d50_K10}
	      \end{subfigure}   
       \vfill
	   \begin{subfigure}{0.52\linewidth}
         \includegraphics[width=\linewidth]{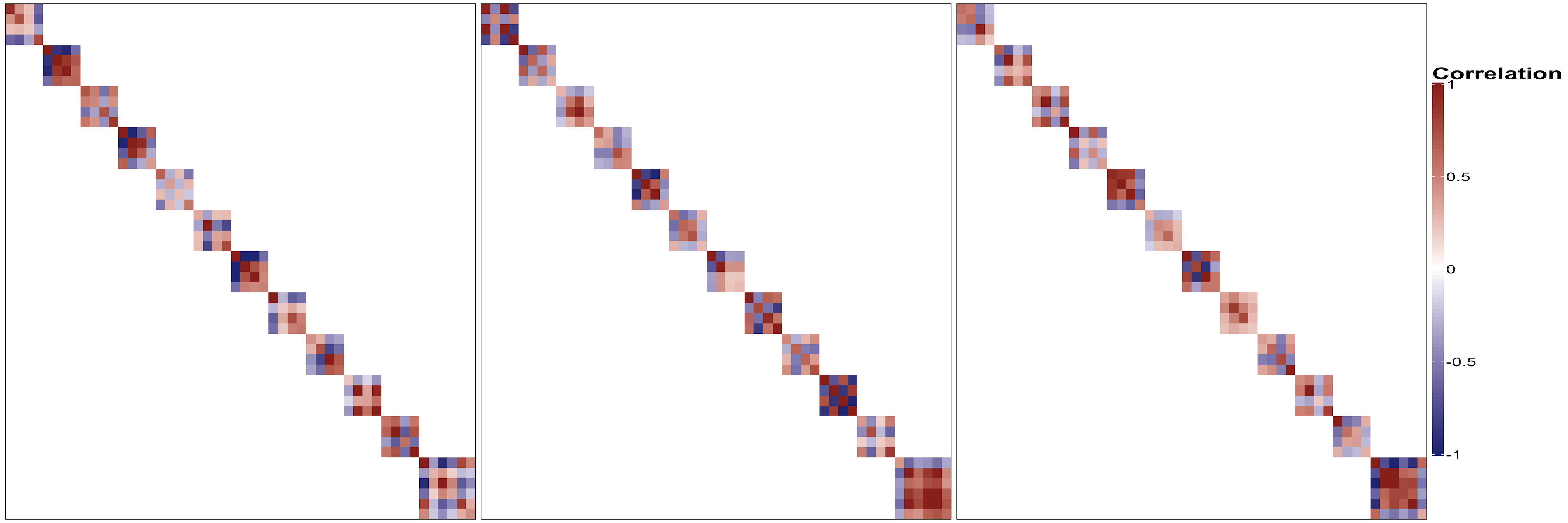}
		 \includegraphics[width=\linewidth]{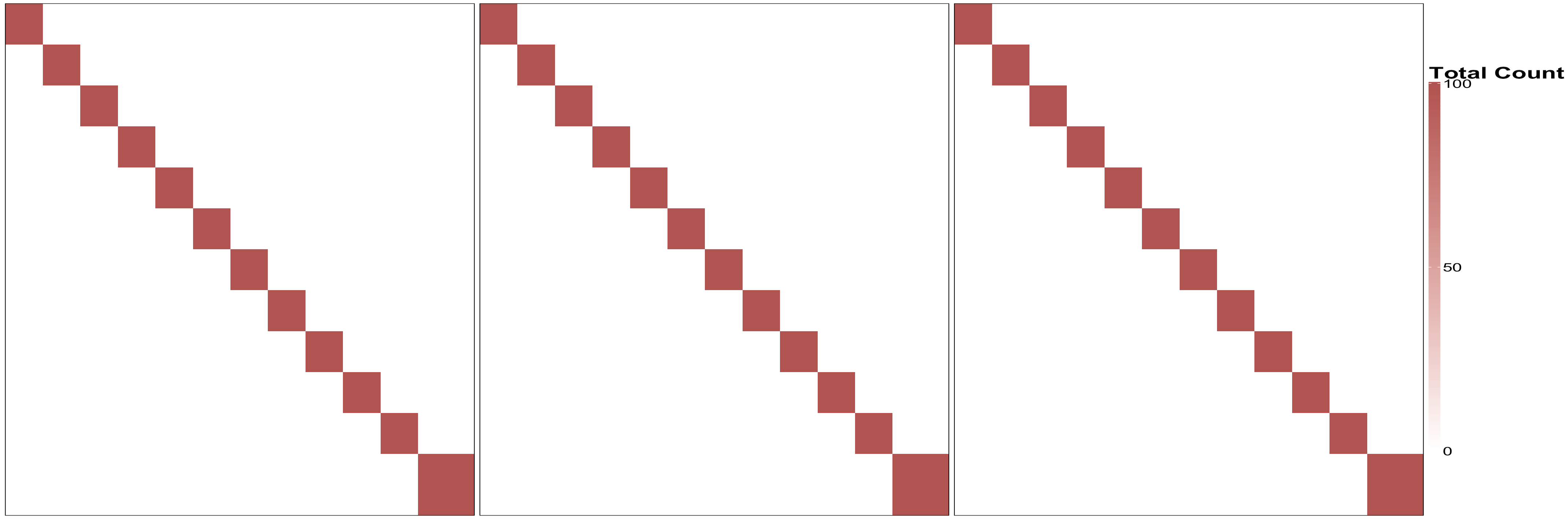}
		  \caption{$N$ = 1000; $d$ = 50; $K$ = 12}
		  \label{fig:d50_K12}
	       \end{subfigure}
	\caption{Covaraince matrices from generated data and the heatmap of the count for a non-zero entry within the covariance matrix for simulations of $N = 1000$ given $d = 50$ with $G = 3$ and $K = 9$; $10$; and $12$ respectively}
\label{fig: N1000 d50 cov matrices1}
\end{figure}
\end{appendices}

\end{document}